\documentclass[twocolumn]{aastex631} 
\usepackage{graphicx}
\usepackage{natbib}
\usepackage{placeins}
\usepackage{amsmath}
\usepackage{tgcursor}
\usepackage{booktabs}
\usepackage{enumitem}
\setcitestyle{notesep={ }}
\usepackage{xcolor}

\usepackage{soul}
\definecolor{newgreen}{rgb}{0., 0.7, 0.4}

\begin{document}

\title{JWST ice band profiles reveal mixed ice compositions in the HH 48 NE disk} 

\author[0000-0002-8716-0482]{Jennifer B.~Bergner}
\affiliation{University of California, Berkeley, Berkeley CA 94720, USA}

\author[0000-0002-0377-1316]{J.~A.~Sturm}
\affiliation{Leiden Observatory, Leiden University, P.O. Box 9513, NL-2300 RA Leiden, The Netherlands}

\author[0000-0001-6947-7411]{Elettra L.~Piacentino}
\affiliation{Center for Astrophysics | Harvard \& Smithsonian, 60 Garden Street, Cambridge, MA 02138, USA}

\author[0000-0003-1878-327X]{M.~K.~McClure}
\affiliation{Leiden Observatory, Leiden University, P.O. Box 9513, NL-2300 RA Leiden, The Netherlands}

\author[0000-0001-8798-1347]{Karin I.~\"Oberg}
\affiliation{Center for Astrophysics | Harvard \& Smithsonian, 60 Garden Street, Cambridge, MA 02138, USA}

\author[0000-0001-9344-0096]{A.~C.~A.~Boogert}
\affil{Institute for Astronomy, University of Hawai’i at Manoa, 2680 Woodlawn Drive, Honolulu, HI 96822, USA}

\author[0000-0003-1197-7143]{E.~Dartois}\affil{Institut des Sciences Mol\'eculaires d’Orsay, CNRS, Univ. Paris-Saclay, 91405 Orsay, France}

\author[0000-0001-7479-4948]{M.~N. Drozdovskaya}
\affiliation{Physikalisch-Meteorologisches Observatorium Davos und Weltstrahlungszentrum (PMOD/WRC), Dorfstrasse 33, CH-7260, Davos Dorf, Switzerland}

\author[0000-0003-0972-1595]{H.~J.~Fraser}
\affiliation{School of Physical Sciences, The Open University, Walton Hall, Milton Keynes, MK76AA, UK}

\author[0000-0001-6307-4195]{Daniel Harsono}
\affiliation{Institute of Astronomy, Department of Physics, National Tsing Hua University, Hsinchu, Taiwan}

\author[0000-0002-2271-1781]{Sergio Ioppolo}
\affiliation{Centre for Interstellar Catalysis, Department of Physics and Astronomy, Aarhus University, DK 8000 Aarhus, Denmark}

\author[0000-0003-1413-1776]{Charles J.\ Law}
\altaffiliation{NASA Hubble Fellowship Program Sagan Fellow}
\affiliation{Department of Astronomy, University of Virginia, Charlottesville, VA 22904, USA}

\author[0000-0002-0500-4700]{Dariusz C.~Lis}
\affil{Jet Propulsion Laboratory, California Institute of Technology, 4800 Oak Grove Dr. Pasadena, CA, 91109, USA}

\author[0000-0003-1254-4817]{Brett A.~McGuire}
\affiliation{Department of Chemistry, Massachusetts Institute of Technology, Cambridge, MA 02139, USA}
\affiliation{National Radio Astronomy Observatory, Charlottesville, VA 22903, USA}

\author[0000-0002-6025-0680]{Gary J.~Melnick}
\affil{Center for Astrophysics | Harvard \& Smithsonian, 60 Garden Street, Cambridge, MA 02138, USA}

\author[0000-0003-4985-8254]{Jennifer A.~Noble}
\affil{Physique des Interactions Ioniques et Mol\'{e}culaires, CNRS, Aix Marseille Univ., 13397 Marseille, France}

\author[0000-0002-9122-491X]{M.~E.~Palumbo}
\affiliation{INAF—Osservatorio Astrofisico di Catania, via Santa Sofia 78, 95123 Catania, Italy}

\author[0000-0001-8102-2903]{Yvonne J.~Pendleton}
\affiliation{Department of Physics, University of Central Florida, Orlando, FL 32816, USA}

\author[0000-0002-8545-6175]{Giulia Perotti}
\affil{Max-Planck-Institut f\"{u}r Astronomie, K\"{o}nigstuhl 17, 69117 Heidelberg, Germany}

\author[0000-0002-3276-4780]{Danna Qasim}
\affiliation{Southwest Research Institute, San Antonio, TX 78238, USA}

\author[0000-0001-6144-4113]{W.~R.~M.~Rocha}
\affiliation{Leiden Observatory, Leiden University, P.O. Box 9513, NL-2300 RA Leiden, The Netherlands}

\author[0000-0001-7591-1907]{E.~F.~van Dishoeck}
\affiliation{Leiden Observatory, Leiden University, P.O. Box 9513, NL-2300 RA Leiden, The Netherlands}
\affil{Max-Planck-Institut für Extraterrestrische Physik, Giessenbachstrasse 1, 85748 Garching, Germany}

\begin{abstract}
\noindent Planet formation is strongly influenced by the composition and distribution of volatiles within protoplanetary disks.  With JWST, it is now possible to obtain direct observational constraints on disk ices, as recently demonstrated by the detection of ice absorption features towards the edge-on HH 48 NE disk as part of the Ice Age Early Release Science program.  Here, we introduce a new radiative transfer modeling framework designed to retrieve the composition and mixing status of disk ices using their band profiles, and apply it to interpret the H$_2$O, CO$_2$, and CO ice bands observed towards the HH 48 NE disk.  We show that the ices are largely present as mixtures, with strong evidence for CO trapping in both H$_2$O and CO$_2$ ice.  The HH 48 NE disk ice composition (pure vs.~polar vs.~apolar fractions) is markedly different from earlier protostellar stages, implying thermal and/or chemical reprocessing during the formation or evolution of the disk.  We infer low ice-phase C/O ratios around 0.1 throughout the disk, and also demonstrate that the mixing and entrapment of disk ices can dramatically affect the radial dependence of the C/O ratio.  It is therefore imperative that realistic disk ice compositions are considered when comparing planetary compositions with potential formation scenarios, which will fortunately be possible for an increasing number of disks with JWST.
\end{abstract}

\section{Introduction}
\label{sec:intro}

Planetary compositions are shaped by the reservoir of volatiles (gas and ice) available during their formation within a protoplanetary disk.  
The bulk elemental ratios (e.g.~C/N/O/H) of a planetary core and primary atmosphere are dependent on the composition of ice and gas accreted during formation.  Additional volatiles and organics may be delivered to planetary surfaces via impacts of icy planetesimals formed in the cold outer disk, contributing chemical building blocks needed for origins of life chemistry \citep[e.g.][]{Marty2012,Rubin2019b}.  While considerable progress has been made in recent years to illuminate the gas-phase compositions in protoplanetary disks \citep[see][and references therein]{Oberg2023}, we still have only a crude understanding of the icy landscape in which planets form.  Indeed, since ices are a major reservoir of volatiles throughout much of the disk, this knowledge gap represents a major limitation in the current understanding of planet formation.  

Observations of protostellar ices show that they typically contain multiple components: pure ices, apolar mixtures, and polar mixtures \citep{Boogert2015}.  Determining whether protostellar-like ice compositions are inherited to the disk stage can provide unique insight into the degree of thermal and/or chemical processing of icy material along the journey to forming planetesimals.  Moreover, the mixing status of disk ices may profoundly impact the distribution of icy volatiles across the disk due to the phenomenon of `entrapment': within H$_2$O- or CO$_2$-dominated ice mixtures, other volatile molecules are expected to remain trapped in the solid phase until temperatures higher than their nominal sublimation temperatures \citep{Collings2004, Simon2019}.  The presence of mixed ices within protoplanetary disks could therefore alter the partitioning of volatile elements between the ice and gas phases compared to expectations for pure ices \citep{Oberg2011b}.

Direct observational constraints of ices in disks are most accessible for sources with an edge-on viewing geometry: the vibrational modes of solid-state ice features, detectable in the near-/mid-IR, are apparent as absorption features against the infrared continuum.  With previous infrared telescopes, absorption features from H$_2$O, CO$_2$, and tentatively CO were detected towards several edge-on disks \citep{Thi2002, Pontoppidan2005, Terada2007, Terada2012, Aikawa2012}.  Signatures of crystalline water were also seen with ISO and \textit{Herschel} \citep{Malfait1998, McClure2015}.  However, these observations suffered from poor sensitivity and a lack of spatial resolution, limiting conclusions about the composition and distribution of ices within the disks.  

Now, with JWST it is possible to obtain high-sensitivity maps of ice features towards edge-on disks, as demonstrated by the JWST Early Release Science Program `Ice Age' \citep{Sturm2023c}.  An initial analysis of this data detected trace ices for the first time in a disk (including NH$_3$, OCS, OCN$^-$, and $^{13}$CO$_2$) and revealed a surprisingly large vertical extent of CO ice absorption.  Moreover, the detection of the major ice species H$_2$O, CO$_2$, and CO at both high signal-to-noise (SNR) and high spectral resolution opens the door to a detailed analysis of the ice band profiles, which is the focus of this work.

Ice band profiles encode information about the `structure' of the ice.  This includes information such as whether the chemical constituents are well-mixed or segregated into distinct layers, and whether the ice itself has an amorphous (disordered) or crystalline (ordered) structure.  This is because the shapes of the ice absorption features are sensitive to the local ice `environment': each of these structural effects impacts the intermolecular interactions, and in turn shifts the vibrational frequencies and alters the band profiles relative to those of an isolated gas-phase molecule.  

With previous generations of infrared facilities, the mixing status of ices in dense clouds and young stellar objects was inferred by matching observed band profiles with experimental templates \citep[e.g.][]{Sandford1988, Tielens1991, Chiar1995, Whittet1998, Dartois1999, Gerakines1999, Pontoppidan2003, Gibb2004, Pontoppidan2008, Perotti2020}.  This method is now being applied to JWST ice observations \citep{Brunken2024}.  
Ice band shapes and intensities are also sensitive to the particle sizes and shapes, and lab spectra are often adjusted to account for these effects prior to comparing with observations \citep{Bohren1983, Tielens1991, Ehrenfreund1997}.  However, even grain shape corrected laboratory spectra cannot be directly used to interpret ice spectra from edge-on disks due to other radiative transfer effects, as prior modeling has demonstrated \citep{Pontoppidan2005,Dartois2022,Sturm2023b}.   

There are several key factors that complicate the radiative transfer through edge-on disks and hinder the interpretation of their spectra.  First, the infrared background itself is a combination of scattered stellar light, thermal emission from warm dust, and scattered emission from warm dust \citep{Pontoppidan2005,Dartois2022, Sturm2023b}.  The balance between scattered light and thermal emission depends on both the grain size ($a$) and the wavelength of light ($\lambda$), with shorter IR wavelengths being more scatter-dominated (scattering becomes important when 2$\pi a / \lambda \sim$ 1).  Due to grain growth to $\gtrsim\mu$m radii, scattering in the IR also plays a more important role in disks compared to earlier stages of star formation.  Scattering can intrinsically alter the ice band profiles, particularly by producing excess absorption or emission in the line wings \citep{Leger1983, Smith1989, Dartois2006, Dartois2022}.  In disks, there is further complication due to the source structure \citep[discussed in detail in][]{Sturm2023b, Sturm2023c}: photons that reach the telescope have likely undergone one or more scattering events and cannot be traced back along a direct path to the star.  Moreover, the escaping photons represent a range of trajectories through the disk; those scattered from the icy midplane may mix with those that pass through the warm disk surface, leading to `dilution' of the apparent ice optical depth.  Altogether, this means that detailed radiative transfer modeling is required to extract information about the quantities and structures of disk ices.

In this paper, we present a new radiative transfer framework designed with the flexibility to reproduce the band profiles of ice features in disks, providing direct constraints on disk ice compositions.  
We apply this model to the NIRSpec \citep{Jakobsen2022} observations of the HH 48 NE disk taken as part of the Ice Age program.  We focus on H$_2$O, CO$_2$, and CO since they are the most abundant ice species in earlier stages of star formation \citep{Boogert2015}; moreover they exhibit strong absorption features in the HH 48 NE near-IR spectrum \citep{Sturm2023c}, which enables detailed analysis of their band profiles.  Section \ref{sec:model} introduces the modeling framework, including the procedure for distributing ice across the disk model and for producing opacities for a range of pure and mixed ices.  In Section \ref{sec:res}, we present simulated H$_2$O, CO$_2$, and CO ice bands for different ice mixing scenarios, and determine the favored ice composition to reproduce the observed spectrum.  We also address the possible presence of CH$_3$OH in the ice, as well as the $^{12}$CO$_2$/$^{13}$CO$_2$ ice abundance ratio.  In Section \ref{sec:discussion}, we interpret the retrieved ice compositions in the context of chemical evolution from protostars to disks to planetesimals, and discuss the C/O ratios implied by our model.  Section \ref{sec:concl} contains our conclusions. 

\FloatBarrier
\section{Modeling framework}
\label{sec:model}

\subsection{Dust + star model}
\label{subsec:dust}

\vspace{-0.15in}
\begin{deluxetable}{lc|lc}
	\tabletypesize{\footnotesize}
	\tablecaption{Properties of the dust disk for HH 48 NE \label{tab:disk_struct}}
	\tablecolumns{4} 
	\tablewidth{\textwidth} 
 	\tablehead{
        \colhead{Parameter}       & 
        \colhead{Value}       & 
        \colhead{Parameter} & 
        \colhead{Value} 
        }
\startdata
$\Sigma_c$ & 1.07 g cm$^{-2}$ & $R_c$ & 87 au \\
$h_c$ & 0.21 & $\psi$ & 0.13 \\
$\gamma$ & 0.81 & $\epsilon$ & 100 \\
$f_\mathrm{lg}$ & 0.89 & $X_\mathrm{lg}$ & 0.2 \\
\enddata
\tablenotetext{}{From \citet{Sturm2023a}}
\end{deluxetable}

We adopt a dust + star model based on \citet{Sturm2023a}, which fitted the properties of HH 48 NE using spectral energy distribution (SED) constraints along with dust and gas images in the millimeter and IR \citep{Stapelfeldt2014, Dunham2016, Villenave2020}.  The structural parameters used for this work are summarized in Table \ref{tab:disk_struct}.  The model is briefly summarized below.  We use a parametric dust model with the radial dust surface density profile described by a tapered power law:
\begin{equation}
    \Sigma_\mathrm{dust}(r) = \frac{\Sigma_c}{\epsilon}\Bigg{(}\frac{r}{R_c}\Bigg{)}^{-\gamma}\; \mathrm{exp}\Bigg{[}-\Bigg{(} \frac{r}{R_c}\Bigg{)}^{2-\gamma} \Bigg{]},
\end{equation}
where $\Sigma_c$ is the surface density at the characteristic radius $R_c$, $\epsilon$ is the gas to dust ratio, and $\gamma$ is the radial power-law index.  The two-dimensional density distribution of the dust is given by:
\begin{equation}
    \rho_\mathrm{dust}(r,\theta) = \frac{\Sigma_\mathrm{dust}}{\sqrt{2\pi}rh} \; \mathrm{exp} \Bigg{[} -\frac{1}{2} \Bigg{(} \frac{(\pi/2 - \theta)}{h}\Bigg{)}^2\Bigg{]},
    \label{eq:rho_dust}
\end{equation}
where $\theta$ is the disk opening angle from the midplane as seen from the position of the central star. $h$ is the scale height, defined by a characteristic scale height ($h_c$) and a power-law index of vertical flaring ($\psi$):
\begin{equation}
    h(r) = h_c \, \Big{(}\frac{r}{R_c} \Big{)}^\psi
\end{equation}
 
The dust is divided into a small and a large population to account for grain growth and settling.  The large grain population accounts for a mass fraction $f_\mathrm{lg}$ of the total dust population.  The small dust distribution follows Equation \ref{eq:rho_dust}, while the large dust vertical extent is described by a large-grain scale height $h_\mathrm{lg} = X_\mathrm{lg} h$.  $f_\mathrm{lg}$ and $X_\mathrm{lg}$ are constants (Table~\ref{tab:disk_struct}).  Both grain populations have a minimum grain size of 0.43 $\mu$m and a power law index of -3.5, with a maximum grain size of 1 $\mu$m for the small population and 1000 $\mu$m for the large population \citep{Sturm2023a}.

For both small and large grains, the dust disk extends radially from an inner sublimation radius of 0.045 au to an outer radius of 300 au.  Following \citet{Sturm2023a}, we include a cavity interior to a 55 au radius, in which the dust density is attenuated by a factor of 1.6$\times$10$^{-2}$.  The star is treated as a blackbody with an effective temperature of 4155 K, a radius of 1.26 R$_\odot$, and a luminosity of 0.4 L$_\odot$.  Based on the dust density and stellar inputs, the dust temperature structure is solved for using the thermal Monte Carlo calculation in \texttt{RADMC-3D} \citep{Dullemond2012} over the wavelength range 10$^{-2}$--10$^4$ $\mu$m with 10$^7$ photons.

It is important to recognize that several of the dust parameters can have an important influence on the modeled ice bands.  This was explored in detail in \citet{Sturm2023b}.  Notably, the degree of dust settling (e.g.~$f_\mathrm{lg}$, $X_\mathrm{lg})$ will affect the vertical location of the optically thick scattering surface.  Also, the grain size distribution affects the amount of ice partitioned onto large grains, which produce less ice absorption relative to small grains.  Because we fix these dust parameters in our modeling, the absolute ice abundances should be treated with some caution, and we focus instead on the ratios between different ice species (Section \ref{sec:discussion}).
 
\subsection{Ice distribution}
\label{subsec:disk_zoning}
Based on the dust and temperature structures described in Section \ref{subsec:dust}, we next map the distribution of ices across the disk.  To account for spatially dependent ice compositions due to the presence of volatile sublimation fronts, we divide the disk into 4 zones based on the freeze-out locations of our target ice species: H$_2$O, CO$_2$, and CO.  Following \citet{Hollenbach2009}, local freeze-out temperatures $T_i$ for each species $i$ are found from:

\begin{equation}
    T_i = E_{b,i} \Bigg{[}57 + \mathrm{ln}\Bigg{(}\frac{1}{n_{\mathrm{H}_2}}\frac{10^4}{c_s}\Bigg{)}\Bigg{]}^{-1}.
    \label{eq:FO}
\end{equation}
$E_{b,i}$ is the binding energy of H$_2$O, CO$_2$, or CO, in each case assuming desorption in the multi-layer regime \citep[Table \ref{tab:zones};][]{Sandford1988b, Noble2012, Fayolle2016, Cuppen2017}.  $n_{\mathrm{H}_2}$ is the gas density.  The sound speed $c_s$ is equal to $(k_b T/\mu_i m_{\text{H}})^{1/2}$, where $k_b$ is the Boltzmann constant, $T$ is the temperature (we adopt the dust temperature assuming equivalent dust and gas temperatures), $\mu_i$ is the molecular weight of the molecule, and $m_{\text{H}}$ is the hydrogen mass.  The midplane snowline locations for each ice species are included in Table \ref{tab:zones}.  

\begin{figure*}
\centering
    \includegraphics[width=\linewidth]{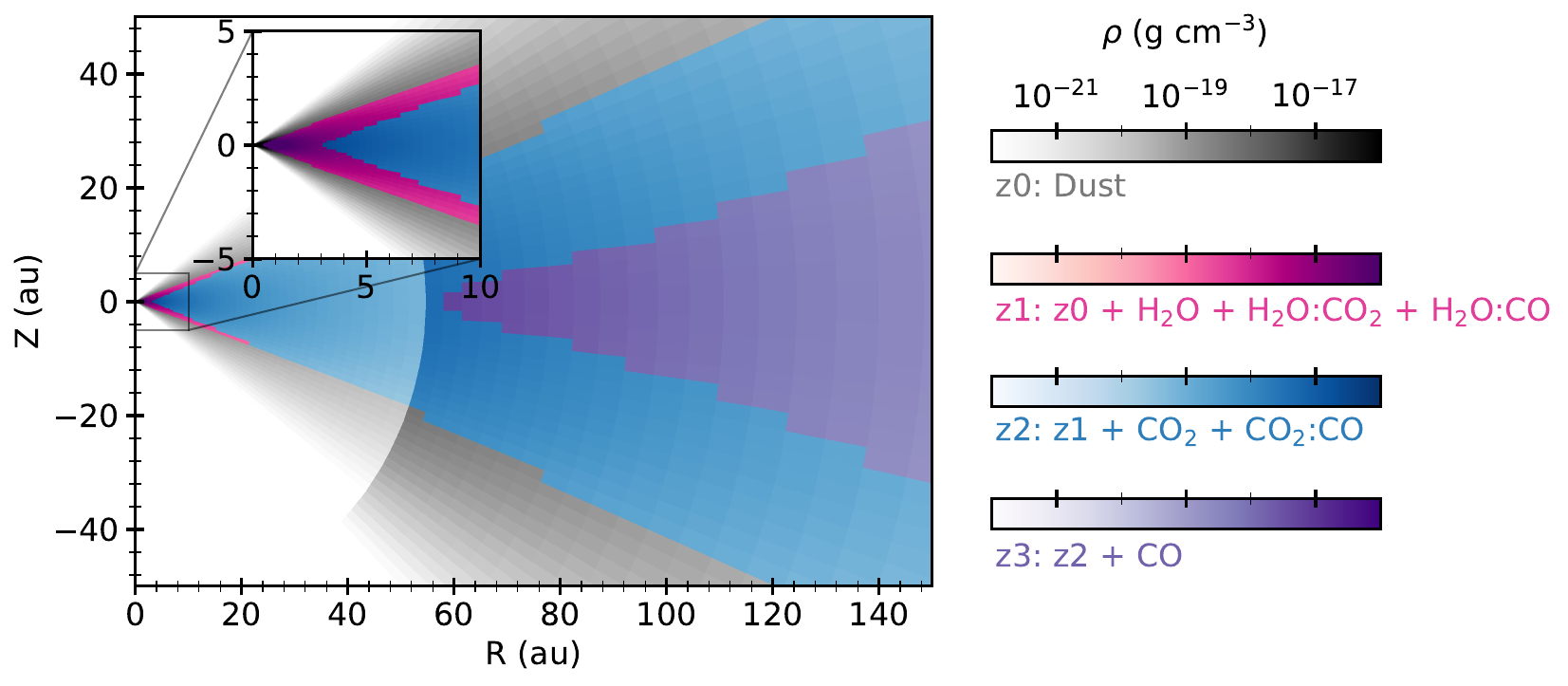}
    \caption{Illustration of disk zoning.  Different colormaps show the density of solids in the small-grain population for each disk zone, where the 4 zones are indicated by z0, z1, z2, and z3.  The allowed solids that may be present within each zone are listed under each colorbar.  Each zone contains new ices along with the solids present in any interior zones (e.g.~zone 1 contains water-dominated ices along with refractory dust from zone 0).  Inset shows a zoom-in of the inner 10 au.}
    \label{fig:zones}
\end{figure*}

By comparing the freeze-out temperatures to the two-dimensional disk temperature structure, we identify the spatial boundary where each volatile will be present in the ice phase.  Note that \citet{Sturm2023b} adopt a single desorption temperature for each species across the disk, resulting in slightly different freeze-out zones.  The resulting zones are labeled 0--3, with zone 0 containing bare dust and zone 3 containing dust along with H$_2$O, CO$_2$, and CO ice.  In addition to these temperature boundaries, we also exclude ices from strongly UV-irradiated upper layers of the disk.  We adopt an ice threshold of A$_V>$ 1.5 with respect to the central star \citep{Sturm2023b} informed by dark cloud observations \citep{Boogert2015}.  Table \ref{tab:zones} summarizes the zone properties, and Figure \ref{fig:zones} illustrates their locations in an example disk model.

If only pure ices were present, there would be a successive condensation of H$_2$O, CO$_2$, and CO moving outwards in the disk.  However, it is well established from experiments that volatiles within ice mixtures can remain trapped in the solid phase at temperatures much higher than their nominal sublimation temperatures.  For instance, in lab experiments, CO can be trapped within a CO$_2$-dominated or H$_2$O-dominated matrix up to temperatures of 80 K and 150 K respectively, compared to a pure ice desorption temperature around 30 K \citep{Collings2004, Fayolle2011, Simon2019}.  To account for this trapping effect, within the zones where H$_2$O is present (zones 1--3) we allow H$_2$O-dominated mixtures to contain CO$_2$ or CO.  Similarly, within the zones where CO$_2$ is present (zones 2--3) we allow CO$_2$-dominated mixtures to contain CO.  The ability to combine different pure and mixed ice mantles within each zone allows us to describe the distribution of ices across the disk as realistically as possible.  Table \ref{tab:zones} summarizes the different ice components that are allowed within each zone.

\vspace{-0.15in}
\begin{deluxetable*}{lcccc}
	\tabletypesize{\footnotesize}
	\tablecaption{Modeled disk zones \label{tab:zones}}
	\tablecolumns{5} 
	\tablewidth{\textwidth} 
	\tablehead{
        \colhead{Zone}       & 
        \colhead{Inner bound$^a$}       & 
        \colhead{$E_b$ (K) $^b$ } & 
        \colhead{Midplane snowline (au)} &
        \colhead{Allowed solids$^c$} 
        }
\startdata
0 & Dust & &  & Refractory grains \\
1 & H$_2$O & 4800 & 0.5 & (0) + H$_2$O + H$_2$O:CO$_2$ + H$_2$O:CO \\
2 & CO$_2$ & 2270 & 3 & (1) + CO$_2$ + CO$_2$:CO \\
3 & CO & 870 & 60 & (2) + CO \\
\enddata
\tablenotetext{a}{Molecule whose sublimation front determines the zone's inner boundary}
\tablenotetext{b}{Binding energy that defines the inner bound; values are all for multi-layer desorption, from \citet{Sandford1988} (H$_2$O), \citet{Noble2012} (CO$_2$), and \citet{Fayolle2016} (CO).}
\tablenotetext{c}{Mixtures of two molecules are denoted A:B}
\end{deluxetable*}

For disk locations where a given pure or mixed ice component is frozen out, its local solid density $\rho_i$ is calculated from: 
\begin{equation}
    \rho_i = \rho_\mathrm{dust} \epsilon \; \frac{X_i \mu_i}{0.64 \mu_\mathrm{gas}}.
    \label{eq:icedens}
\end{equation} 
$X_i$ represents the fractional abundance of the molecule with respect to H, and we assume local gas densities of $\rho_\mathrm{dust} \epsilon$, where $\epsilon$ is the gas to dust mass ratio.  $\mu_i$ is the mean molecular weight of the ice component.  $\mu_\mathrm{gas}$ is the gas mean molecular weight, taken here to be 2.37 $m_\mathrm{H}$, and the fraction 0.64 corresponds to the gas abundance (composed of H$_2$ and He).  $\rho_i$ is then apportioned between small and large grains: the ice density assigned to large grains is found from scaling $\rho_i$ by the fraction $f_{i,l}$, and the ice density assigned to small grains is then $\rho_i(1-f_{i,l})$.  $f_{i,l}$ is determined based on the relative surface area of the large vs.~small grains, following \citet[][]{Ballering2021}:

\begin{equation}
    f_{i,l} = f_\mathrm{lg}\Bigg{[}\frac{a_\mathrm{max, l}}{a_\mathrm{max, s}} \;\Big{(}1-f_\mathrm{lg}\Big{)}+f_\mathrm{lg} \Bigg{]}^{-1/2}.
    \label{eq:f_ice}
\end{equation}
Again, $f_\mathrm{lg}$ is the mass fraction of dust in large grains, and the maximum grain sizes of the small and large grain populations are represented with $a_\mathrm{max, s}$ and $a_\mathrm{max, l}$, respectively.  The total density of solids at a given disk position is the sum of dust and any ices present.  For each small and large dust population within each zone, opacity spectra are generated based on the average composition of solids across the zone, as described in Section \ref{subsec:opt_calc}.

\subsection{Ice spectra and optical constants}
\label{subsec:nk_lab}
In order to flexibly fit the observed profiles of the near-IR H$_2$O, CO$_2$, and CO ice bands, we required a reference library of experimentally-derived ice spectra for different mixing ratios.  While there are numerous IR spectra of such ice mixtures available from previous studies, we chose to re-measure the required spectra in order to minimize systematics arising from setup-to-setup variations in e.g.~temperature calibration, ice porosity, and sample alignment.  This ensures that any changes in the experimental band profiles are unambiguously due to ice mixing status.

\begin{figure*}
\centering
    \includegraphics[width=\linewidth]{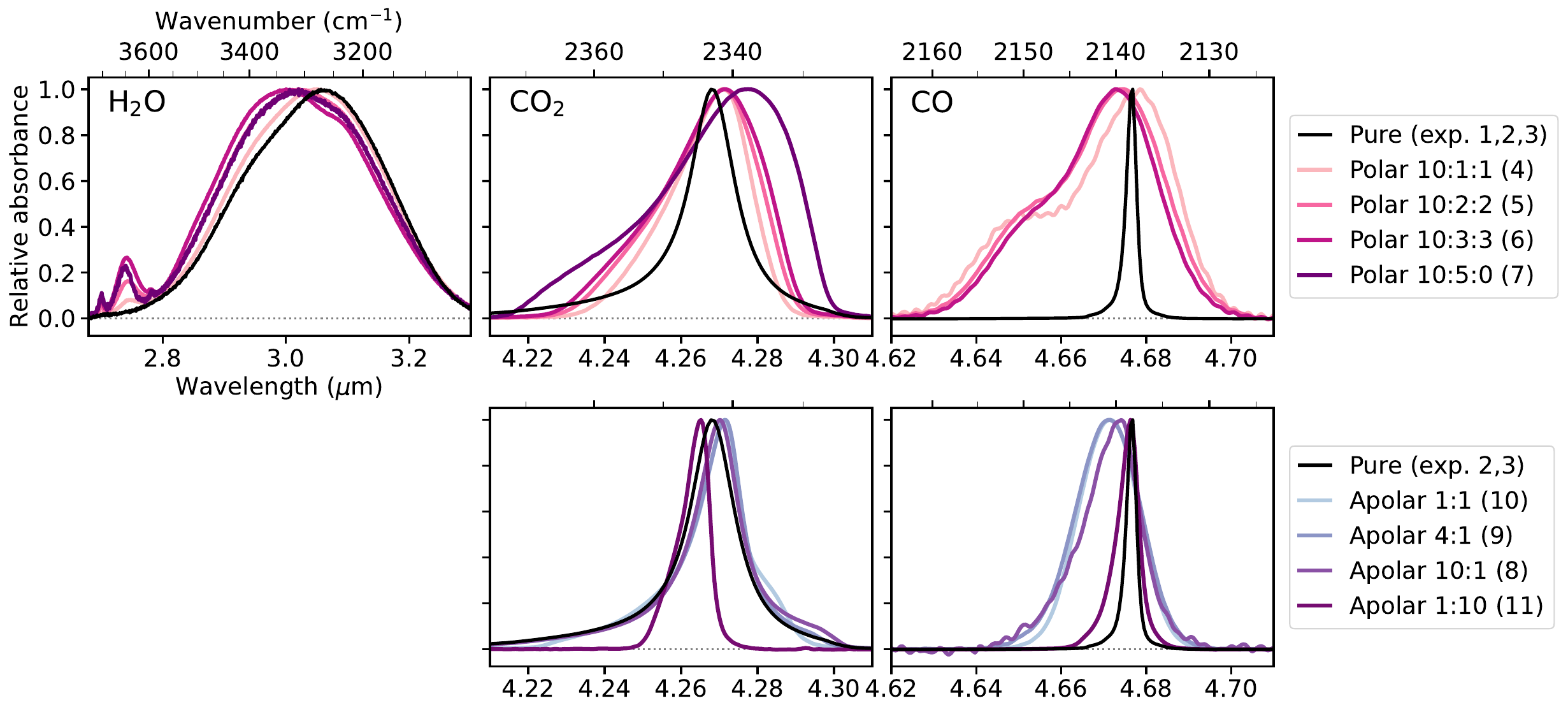}
    \caption{Lab-measured absorbance spectra of different H$_2$O, CO$_2$, and CO ice mixtures, normalized to show differences in the band profiles.  Polar mixtures are shown in the top row, and apolar mixtures in the bottom row.  See Table \ref{tab:exps} for more details on the mixing ratios.  }
    \label{fig:exp_spec}
\end{figure*}

We are primarily focused on the influence that the major ice species (H$_2$O, CO$_2$, and CO) have on each others' near-IR bands.  To this end, we measured a series of spectra with systematically varying ice mixing ratios: pure ices; polar mixtures with H$_2$O:CO$_2$:CO ratios ranging from 10:1:1 to 10:3:3; and apolar mixtures with CO$_2$:CO ratios ranging from 10:1 to 1:10 (Table \ref{tab:exps}).  Note that the polar mixtures also include CH$_4$ at a $<$5\% level, which does not meaningfully affect the H$_2$O, CO$_2$, or CO band profiles.  The experimental spectra are shown in Figure \ref{fig:exp_spec}.  We also include several mixtures containing CH$_3$OH, as it is inferred to influence the band profiles of CO$_2$ and CO in other star-forming stages \citep{Ehrenfreund1998, Cuppen2011}.

\vspace{-0.15in}
\begin{deluxetable}{lccccc}
	\tabletypesize{\footnotesize}
	\tablecaption{Summary of experiments \label{tab:exps}}
	\tablecolumns{6} 
	\tablewidth{\textwidth} 
	\tablehead{
        \colhead{\#}       & 
        \colhead{Name} & 
        \multicolumn{4}{c}{Column densities (10$^{15}$ cm$^{-2}$)} \\[-0.1cm]
        \colhead{} &
        \colhead{} &
        \colhead{H$_2$O} &
        \colhead{CO$_2$} &
        \colhead{CO} & 
        \colhead{CH$_3$OH}} 
\startdata
1 & Pure H$_2$O & 111 & 0   & 0   & 0 \\
2 & Pure CO$_2$ & 0   & 122 & 0   & 0 \\
3 & Pure CO     & 0   & 0   & 130 & 0 \\
4 & Polar 10:1:1$^a$ & 306 & 30 & 7.1 & 0 \\
5 & Polar 10:2:2$^a$ & 264 & 56 & 46 & 0 \\
6 & Polar 10:3:3$^a$ & 224 & 83 & 69 & 0 \\
7 & Polar 10:5:0 & 66 & 32 & 0 & 0 \\
8 & Apolar 10:1 & 0 & 79 & 6.9 & 0 \\
9 & Apolar 4:1 & 0 & 167 & 43 & 0 \\
10 & Apolar 1:1 & 0 & 114 & 104 & 0 \\
11 & Apolar 1:10 & 0 & 8.5 & 79 & 0 \\
\hline
12 & CO$_2$:CH$_3$OH 1:1 & 0 & 41 & 0 & 38 \\
13 & CO:CH$_3$OH 1:1 & 0 & 0 & 52 & 75 \\
\enddata
\tablenotetext{}{Spectra used in this work were obtained at 20 K or 70 K.}
\tablenotetext{a}{These mixtures also include CH$_4$ at a $<$5\% level.}
\end{deluxetable}

Spectra were obtained using the ultra-high vacuum setup SPACEKITTEN, described in detail in \citet{Simon2023}.  Briefly, ices were grown by introducing gas vapor into the chamber through a dosing tube, where it is condensed onto a cryogenically cooled CsI substrate held at 12 K.  All dosing is performed with the dosing tube at a normal angle of incidence to the substrate.  For multi-component ices, the vapors were pre-mixed in the gas mixing line prior to dosing, with final mixing ratios determined from the IR spectrum.  A Bruker Vertex 70v Fourier transform IR spectrometer was used to obtain absorbance spectra from 600--4000 cm$^{-1}$ (17--2.5 $\mu$m) with a resolution of 0.5 cm$^{-1}$.  In the near-IR range ($<$5$\mu$m), this corresponds to a resolution R/$\Delta$R $\sim$4000-8000, compared to 2700 for the JWST NIRSpec spectrum \citep{Sturm2023c}.  Once deposited, ices were warmed at a constant rate of 2 K/min, with spectra obtained in 10 K increments until complete sublimation of the ice sample occurred, which ranges from $\sim$40--150 K depending on the ice composition.  All experimental spectra are available at Zenodo \citep{zenodo}.  For this work, we mainly used spectra obtained at 20 K since pure CO begins to sublimate at higher temperatures, and warmer H$_2$O and CO$_2$ ice spectra (40 K) were found to provide worse matches to the observed band profiles.  We also used several 70 K spectra to test a higher-temperature ice component, which corresponds to the highest-temperature spectra prior to CO$_2$ sublimation (Appendix \ref{subsec:app_co2wing}).

For a given experiment, the column density $N_i$ of each ice species was determined from the IR bands using:
\begin{equation}
    N_i = \frac{\smallint{\tau_i}d\Tilde{\nu} }{A_i},
    \label{eq:columns}
\end{equation}
where $\int{\tau_i}d\Tilde{\nu}$ is the integrated optical depth over the target band and $A_i$ is the band strength.  Band strength uncertainties are the major source of uncertainty on the measured column densities, typically taken to be $\sim$20\% \citep[e.g.][]{Bouilloud2015}.  The total thickness of the ice ($t$), which is needed to derive optical constants, could then be found from the sum of individual ice column densities:

\begin{equation}
    t = \Sigma_i \frac{N_i m_i}{\rho_i},
    \label{eq:thickness}
\end{equation}
where for each species, $m_i$ is the molecular weight and $\rho_i$ is the density.  For a given species, we adopt a constant $\rho_i$ for either a pure or mixed environment due to a lack of mixture-specific density measurements; this may introduce errors on a $\lesssim$10\% level to the calculated thicknesses.  Table \ref{tab:lab_constants} lists the adopted band strengths and ice densities.  

\vspace{-0.15in}
\begin{deluxetable}{lcccc}
	\tabletypesize{\footnotesize}
	\tablecaption{Ice properties \label{tab:lab_constants}}
	\tablecolumns{5} 
	\tablewidth{\textwidth} 
	\tablehead{
        \colhead{Molecule}       & 
        \colhead{Band position} & 
        \colhead{Band strength} & 
        \colhead{Density} &
        \colhead{$n_\mathrm{vis}$} \\[-0.1cm]
        \colhead{} &
        \colhead{(cm$^{-1}$)} &
        \colhead{(cm molec.$^{-1}$)} &
        \colhead{(g cm$^{-3}$)} & 
        \colhead{}
        }
\startdata
H$_2$O & 3280 & 1.5$\times$10$^{-16}$  & 0.87 & 1.27 \\
CO$_2$ & 2340 & 7.6$\times$10$^{-16}$  & 1.11 & 1.27 \\
CO     & 2140 & 1.12$\times$10$^{-17}$ & 0.80 & 1.25 \\
CH$_3$OH & 1031 & 1.07$\times$10$^{-17}$ & 1.01 & 1.33 \\
\enddata
\tablenotetext{}{Values are from the measurements of \citet{Bouilloud2015}.}
\end{deluxetable} 

To provide even more flexibility in fitting the observed spectra, we performed a two-dimensional interpolation of the absorbance of each ice band as a function of both wavelength and mixing ratio.  This then allows us to generate a synthetic IR spectrum for any arbitrary mixing ratio of H$_2$O:CO$_2$:CO ice within the measured mixing ranges.  The interpolation routine is described in detail in Appendix \ref{sec:app_interp}.

For each ice spectrum, we next extracted the wavelength-dependent optical constants $n$ and $k$, corresponding to the real and imaginary parts of the complex refractive index.  Following a linear local baseline subtraction around each feature, $n$ and $k$ were calculated for each absorption spectrum using \texttt{nkice} \citep{Gerakines2020}.  For H$_2$O, we have verified that the optical constants calculated for the 3$\mu$m band are nearly identical regardless of whether the 6.5 and 11$\mu$m features are included in the calculation.  We use the CsI optical constants reported in \citet{Li1976}.  We adopt the refractive indices measured at visible wavelengths ($n_\mathrm{vis}$) from \citet{Bouilloud2015} for each molecule, listed in Table \ref{tab:lab_constants}.  Uncertainties in the refractive index measurements may influence the simulated ice features at a level of $\sim$5-20\% \citep{Rocha2024}.

\begin{figure}
\centering
    \includegraphics[width=\linewidth]{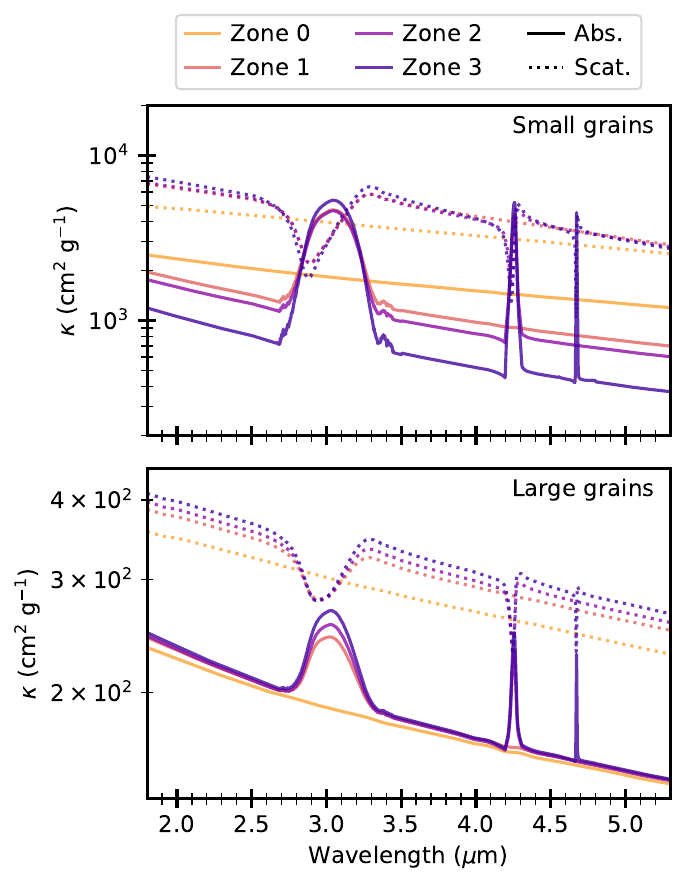}
    \caption{Absorption and scattering opacities for the small- and large-grain populations within each of the four disk zones.  Model M1 is used for illustration (Table \ref{tab:abundances}).}
    \label{fig:opacities}
\end{figure}

\subsection{Dust \& ice opacities}
\label{subsec:opt_calc}
We used \texttt{OpTool} \citep{Dominik2021} to calculate solid-phase opacities for the small and large grain populations in each disk zone (Section \ref{subsec:disk_zoning}).  Following \citet{Sturm2023a}, we assume the refractory grain core has a mass fraction of 85\% amorphous pyroxene (Mg$_{0.8}$Fe$_{0.2}$SiO$_3$) and 15\% amorphous carbon.  Opacities are calculated over the grain size distributions described in Section \ref{subsec:dust} using 50 size bins.  We used the \texttt{-chop} switch to avoid artifacts in the radiative transfer due to extreme forward-scattering, capping the first 3$^\circ$ of the forward-scattering peak.  Both the core and mantle are assumed to have a 25\% porosity \citep[e.g.][]{Kaeufer2023}.

\begin{figure*}
\centering
    \includegraphics[width=\linewidth]{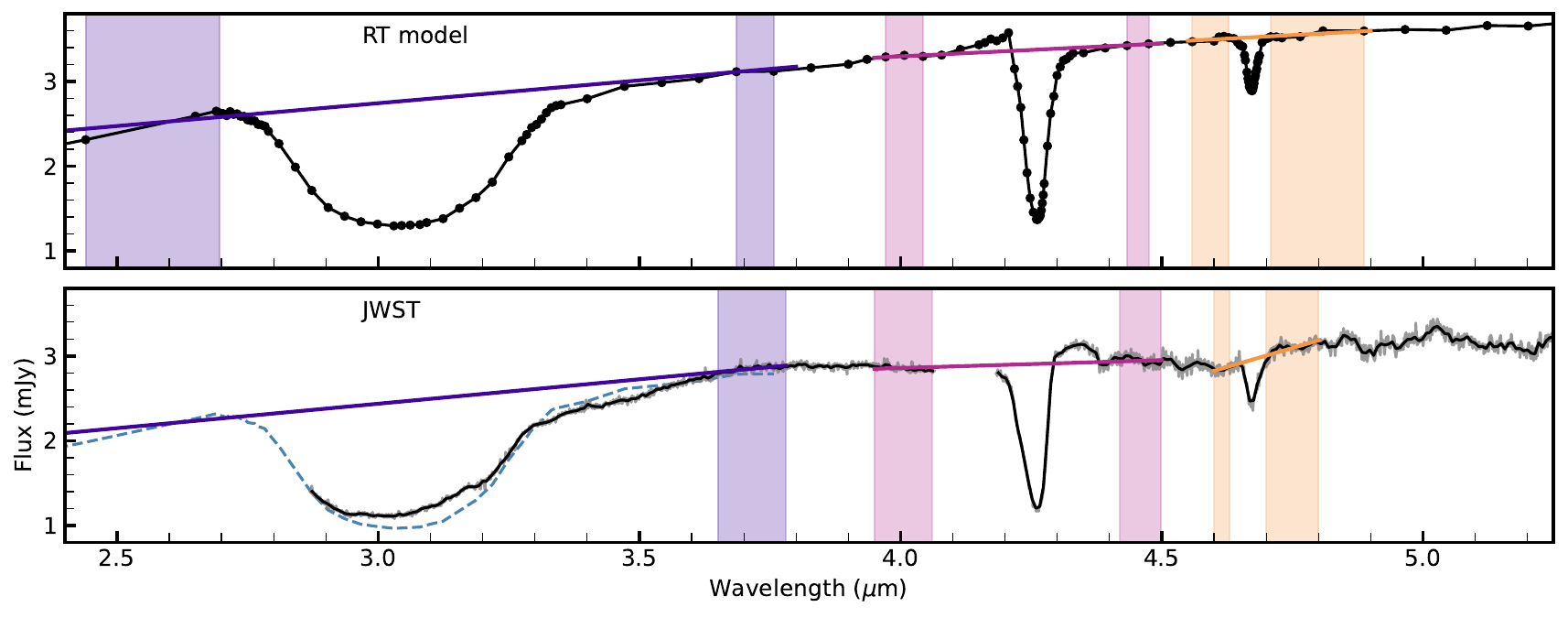}
    \caption{Top: Full near-IR spectrum from a radiative transfer model, including the wavelength ranges used to determine the local baselines for H$_2$O (purple), CO$_2$ (pink), and CO (orange).  Circle markers show the wavelengths sampled for ray tracing.    Bottom:  As above, but for the observed JWST spectrum \citep{Sturm2023c}.  The dashed blue line shows the radiative transfer model (with offset) used to predict the continuum around the H$_2$O band.  4.08–4.19$\mu$m is not covered by the observations.}
    \label{fig:baselines}
\end{figure*}

For pyroxene and carbon, refractive indices used in the \texttt{OpTool} calculations correspond to the measurements of \citet{Dorschner1995} and \citet{Zubko1996}, respectively.  For ice species, we use the lab-derived optical constants described in Section \ref{subsec:nk_lab}.  Note that the lab measurements cover only the range from $\sim$2.5--15 $\mu$m, so the grain opacities outside of this range are determined entirely from the refractory grain component.  This does not impact the resulting near-IR ice absorption spectra. 

In order to capture the effects of a population of varying particle shapes, we use the distribution of hollow spheres (DHS) method with a maximum hollow volume fraction of 0.8.  This method accounts for the optical properties of irregular particle ensembles \citep{Min2005}.  Note that this assumes monomer grains, when in reality the large-grain population may be better described as aggregates.  While further work will be needed to explore the impact of aggregate grains within our framework, detailed modeling has shown that the mass opacity spectra of monomer and aggregate grain distributions tend to be similar, with grain size distribution playing a much more important role in influencing the optical properties \citep{Dartois2022}.

Within \texttt{OpTool}, two grain components may be specified: a `core' and `mantle'.  These two components are added together using the Maxwell-Garnett rule, in which they are treated as a mixture with the core being an inclusion within the mantle \citep{Bohren1983}.  The Bruggeman rule, which treats different mixture components interchangeably, is used to account for the mixing of multiple materials within either the core or mantle.  We include refractory materials (pyroxene and carbon) in the `core' component and ice mixtures within the `mantle' component.  For both the small- and large-grain populations within each disk zone, we produce a single set of opacities based on the average mass fraction of the icy and refractory solid components across the zone.  Example opacities are shown in Figure \ref{fig:opacities}.

\subsection{Comparison to observations}
\label{subsec:method_compare}
With the stellar inputs and spatial distribution of solids from Sections \ref{subsec:dust} and \ref{subsec:disk_zoning}, and opacities from Section \ref{subsec:opt_calc}, we then produce synthetic image cubes of the HH 48 NE disk using the anisotropic scattering treatment in \texttt{RADMC-3D}.  We adopt a viewing inclination of 82.3$^\circ$ \citep{Sturm2023a}.  6$\times$10$^6$ photons are used for the scattering Monte Carlo calculation in order to produce spectra with sufficiently low noise for band profile comparisons.  Images are generated for 144 wavelengths spanning 1.6--5.28$\mu$m, with wavelength sampling chosen by-hand to capture the band profiles of the H$_2$O, CO$_2$, and CO bands at 3, 4.27, and 4.67 $\mu$m respectively ($\lambda$/$\Delta \lambda \sim$100--2000 over the lines).  To compare to the NIRSpec observations of HH 48 NE, we produce a disk-integrated spectrum and apply an extinction correction based on the extinction law in \citet{McClure2009} and assuming an A$_V$ of 5 due to foreground clouds \citep{Sturm2023a}.  We then fit a linear continuum baseline around the ice bands of interest (Figure \ref{fig:baselines} top) and retrieve spectra in optical depth units.  Note that for both CO$_2$ and CO, the baseline regions were chosen to avoid including scattering wings in the continuum fit.

\begin{figure*}
\centering
    \includegraphics[width=\linewidth]{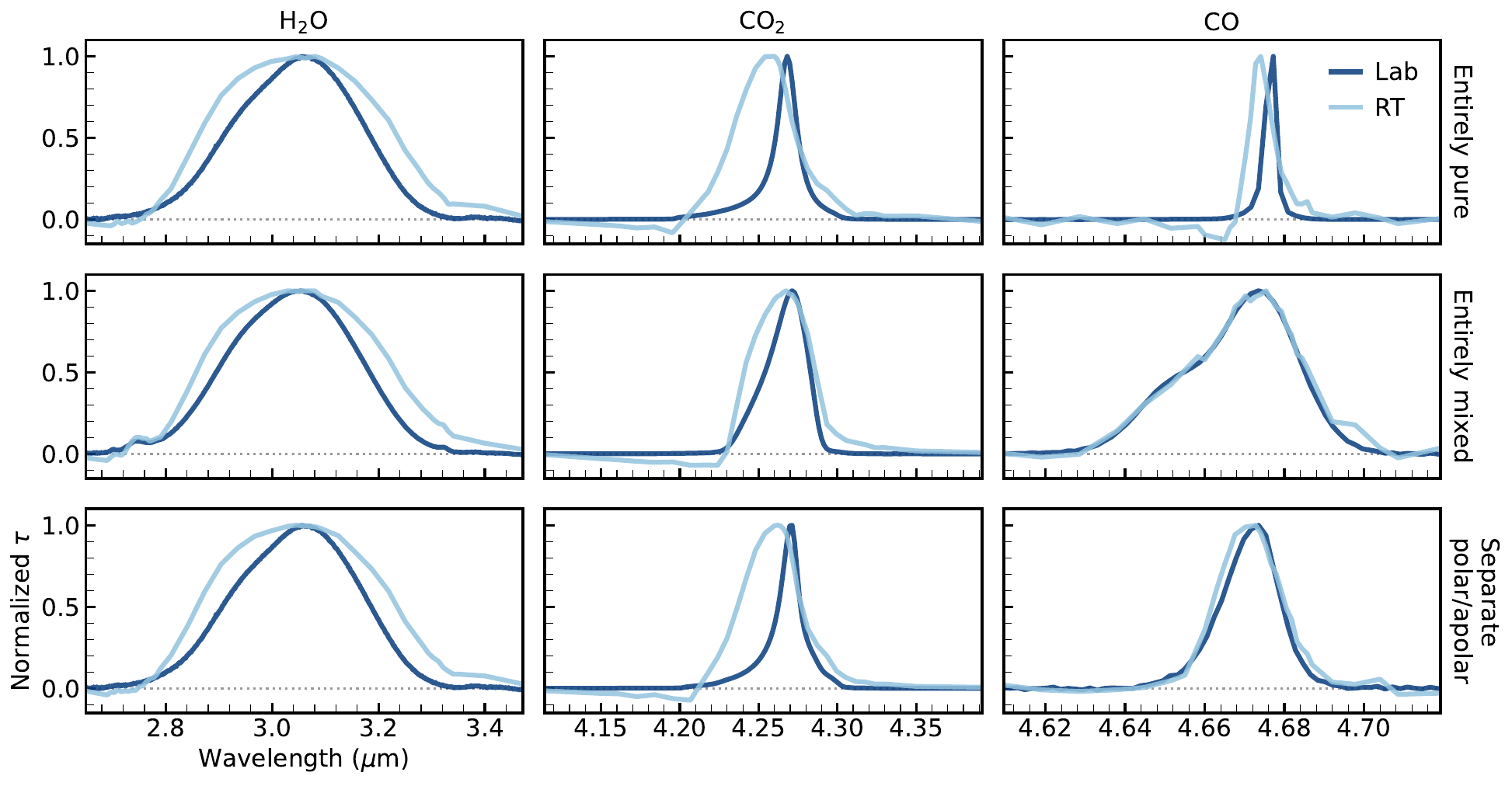}
    \caption{Experimentally measured ice optical depth profiles (dark blue) compared to the band profiles following radiative transfer modeling (light blue).  Spectra are shown for the H$_2$O, CO$_2$, and CO bands and for each of the single-component mixing scenarios described in Section \ref{subsec:RT_v_exp}.}
    \label{fig:exp_vs_RT}
\end{figure*}

For the JWST spectrum (presented originally in \citealt{Sturm2023c}), we similarly fit local linear baselines around the ice bands to extract optical depth spectra.  The continuum subtraction is more challenging for the observations due to gaps in wavelength coverage as well as noisier data, especially around the CO band.  For the H$_2$O band, we are lacking continuum coverage on the blue side.  As shown in Figure \ref{fig:baselines} (bottom), using only the red-side continuum, we chose a linear baseline which reproduces the expected blue continuum based on our radiative transfer models.  Note that spectral coverage below 2.8 $\mu$m is needed to confirm this baseline choice.  For CO$_2$, although the NIRSpec gap coincides with the blue wing, we are still able to fit to the continuum at lower wavelengths.  By testing other possible baseline choices, we estimate that the uncertainties in our peak optical depths introduced by the choice of baseline are $\sim$15\%, 5\%, and 10\% for H$_2$O, CO$_2$, and CO respectively.  The line profiles are not strongly affected, and our main conclusions on ice mixing should be insensitive to the choice of baseline.

\section{Results}
\label{sec:res}

With the framework described in Section \ref{sec:model}, we are able to tune the ice abundances and compositions to reproduce the optical depths and profiles of the H$_2$O, CO$_2$, and CO ice bands observed towards HH 48 NE.  Given the computational expense of the models, it is currently not practical to perform a statistical fitting of the data with our model, and we instead tune the inputs by hand.  In Section \ref{subsec:RT_v_exp}, we first illustrate how the input laboratory spectra are altered by radiative transfer through the disk, adopting simple edge cases for ice mixing (pure, mixed, or separate polar/apolar components).  We then attempt to reproduce the observations using these ice mixing scenarios, subsequently termed `single-component' models, in Section \ref{subsec:res_endmemb}.  Upon showing that none of these scenarios can reproduce the observed ice band profiles, in Section \ref{subsec:res_finetune} we allow for combinations of mixed and pure ices in the model, subsequently termed `multi-component' models.  Our goal is to provide a single description of the disk ice composition that can simultaneously reproduce the H$_2$O, CO$_2$, and CO ice band depths and profiles, and we therefore consider all three bands any time we compare a model to the observations.  The ice abundances used in the all models presented here can be found in Appendix \ref{sec:app_abund}, along with an estimate of the abundance uncertainties.

\subsection{Impact of radiative transfer on ice band profiles \label{subsec:RT_v_exp}}
We begin by considering the effect of radiative transfer on the ice band profiles observed in the HH 48 NE disk.  For this comparison, we tested the following simple edge cases for ice mixing:
\begin{enumerate}[nolistsep]
    \item Entirely pure ices: H$_2$O, CO$_2$, and CO exist as separate, non-interacting ice layers.
    \item Entirely mixed ices: H$_2$O, CO$_2$, and CO are all co-mixed and interacting.
    \item Separate polar and apolar ices: CO$_2$ and CO are co-mixed but are present in a separate layer from H$_2$O.
\end{enumerate}

Figure \ref{fig:exp_vs_RT} shows the optical depth profiles from the experimental spectra (i.e.~before radiative transfer) compared to the profiles after radiative transfer.  For H$_2$O, differences in the initial experimental spectra are essentially erased after radiative transfer.  This is attributable to saturation filling in the line wings.  While the apparent optical depth of H$_2$O in HH 48 NE is less than 1, it suffers from `local saturation' as described in \citet{Sturm2023b} and \citet{Sturm2023c}.  In essence, this is due to the mixing of light scattered from optically thick parts of the disk together with unattenuated light, producing a lower apparent optical depth.

An important consequence is that it is impossible to infer the mixing status of the H$_2$O ice from the band shape at wavelengths above $\sim$2.8 $\mu$m.  The 2.75$\mu$m `dangling' OH feature \citep{Buch1991, McCoustra1996, He2018} does remain diagnostic of H$_2$O mixing even after radiative transfer, appearing only when H$_2$O is mixed with other volatiles (Figure \ref{fig:exp_vs_RT} middle row).  While this feature is well characterized in experimental spectra, it was only recently detected for the first time in astrophysical spectra thanks to JWST \citep{Noble2024}.  Spectral coverage below 2.7 $\mu$m will therefore be critical for future efforts to determine the H$_2$O mixing status.

\begin{figure*}
\centering
    \includegraphics[width=\linewidth]{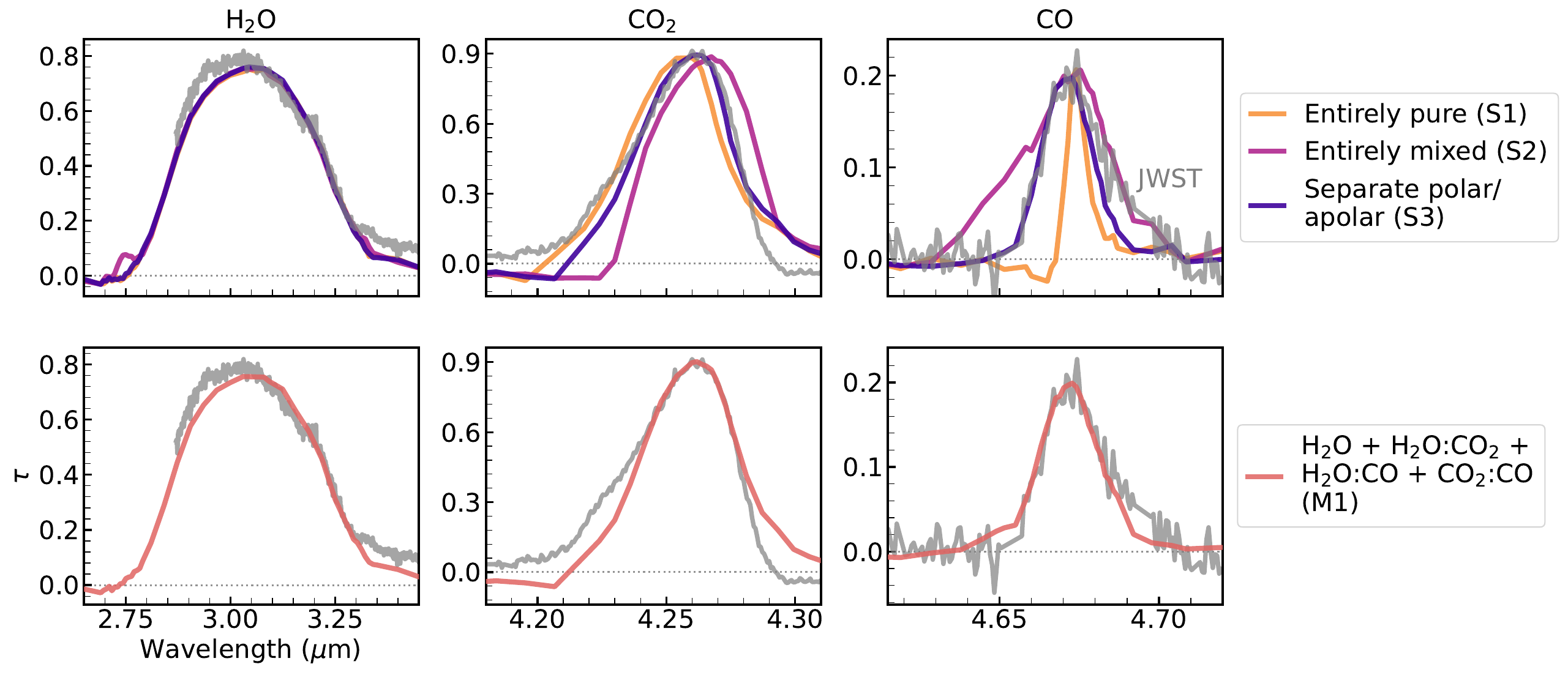}
    \caption{Simulated HH 48 NE ice band profiles (colors) compared to the spectrum observed with JWST (grey).  Top: models of simple ice mixing scenarios in which the ices are assumed to be entirely pure (non-interacting), entirely mixed (interacting), or to have a polar (H$_2$O) layer separate from a mixed apolar (CO$_2$:CO) layer.  Bottom: the preferred multi-component model, including both polar and apolar components of CO$_2$ and CO.  Ice abundances for each model are listed in Table \ref{tab:abundances} with the model names in parentheses.}
    \label{fig:sim_spec}
\end{figure*}

The CO$_2$ band profile is also affected by radiative transfer, though less so than for H$_2$O.  The different mixing scenarios do produce distinguishable spectra, though the differences are smaller than in the experimental spectra.  The line profile changes are in part attributable to local saturation filling in the line wings, as described for H$_2$O.  The CO$_2$ wings are additionally quite sensitive to grain scattering effects \citep{Dartois2022}, which in our models produce a slight blue absorption deficit and red absorption excess around the CO$_2$ band.

The CO band shapes are not as affected by radiative transfer effects compared to CO$_2$ and H$_2$O, and are indeed quite similar to the initial laboratory spectra.  The main effect is for the pure-CO case (Figure \ref{fig:exp_vs_RT} top row), which exhibits a slight blue-ward peak shift in addition to scattering wings.  This is likely due to the narrowness of the feature, combined with the fact that the disk region with pure CO freeze-out has a high fraction of large grains \citep{Dartois2022}.

It is clear that, particularly for H$_2$O and CO$_2$, experimentally measured ice spectra cannot be directly compared to observations of disk ice spectra due to changes in the band profiles induced by radiative transfer.  Still, even after radiative transfer the band profiles of CO$_2$ and CO remain sensitive to the ice mixing status, confirming that mixing status can be inferred using detailed modeling.

\subsection{Single-component ice models}
\label{subsec:res_endmemb}
We first attempted to fit the HH 48 NE spectra using the simple edge case scenarios for ice mixing described in Section \ref{subsec:RT_v_exp}.  Figure \ref{fig:sim_spec} (top) shows the observed NIRSpec H$_2$O, CO$_2$, and CO ice band optical depths compared to these scenarios.  The modeled abundances have been chosen to match the peak optical depths of the observed H$_2$O, CO$_2$, and CO bands, and are reported in Table \ref{tab:abundances}. 

As expected from Section \ref{subsec:RT_v_exp}, the different models do not produce distinguishable H$_2$O spectra when considering only the observed portion of the spectrum $>$2.87 $\mu$m.  Nonetheless, all three scenarios can reasonably reproduce the observed H$_2$O band profile, with the excess absorption in the observed spectrum possibly due to NH$_3$ (2.93 $\mu$m), crystalline H$_2$O or PAH emission (3.2 $\mu$m), and hydrates (3.5 $\mu$m) \citep{Sturm2023c}.

For CO$_2$, none of the simple mixing scenarios can fully reproduce the shape of the observed CO$_2$ band: the entirely-pure ice model (S1) has excess absorption on the blue wing, while the entirely-mixed ice model (S2) has excess absorption on the red wing.  The CO$_2$:CO mixture (S3) is closest to the observations, though does not match the shape at the line peak.  Note that the observed CO$_2$ band also exhibits distinct scattering wing, which appear as an absorption excess on the blue-side baseline and an absorption deficit on the red-side baseline.  This is not reproduced by any of our models, which actually show a small blue deficit and red excess.  These CO$_2$ scattering wings are influenced by the grain size and shape distributions \citep{Dartois2022}, as well as the scattering angles within the disk \citep{Sturm2024}.  The grain size distribution of our model is fixed by SED fitting \citep{Sturm2023a}, and we opt not to change this parameter when attempting to reproduce the CO$_2$ wings.  We focus our comparisons on the core of the CO$_2$ band ($\sim$4.24--4.28 $\mu$m), which should be least impacted by scattering effects.  However, a more self-consistent treatment of both ice composition and grain scattering physics will be needed to confirm our inferences based on the CO$_2$ band profile.

For CO, the different mixing scenarios are clearly distinguishable, with the band profile increasing in width from pure CO (S1) to CO:CO$_2$ (S2) to CO:H$_2$O (S3).  The observed band is too broad to be entirely due to pure CO, and too narrow to be entirely CO in a H$_2$O matrix.  The CO:CO$_2$ mixture provides a good fit to the blue wing but underproduces the red wing of the observed spectrum.  These results imply that the majority of CO exists in a mixture with CO$_2$, with a smaller component present in a polar matrix. 

Based on these results, it is clear that none of our single-component ice models can reproduce the observed band profiles.  We next consider that multiple pure and mixed ices may be present in the disk.

\subsection{Multi-component ice models}
\label{subsec:res_finetune}

Of the single-component models, the separate polar/apolar composition provides the closest match to the observed CO$_2$ and CO band profiles.  By adding a small amount of polar H$_2$O:CO$_2$ and H$_2$O:CO mixtures, we are able to provide a better match to both the CO$_2$ and CO ice bands.  Figure \ref{fig:sim_spec} (bottom) shows the resulting spectrum for an ice model consisting of H$_2$O, 10:1 H$_2$O:CO$_2$, 10:1 H$_2$O:CO, and 3.5:1 CO$_2$:CO.  A detailed comparison with other multi-component ice compositions is presented in Figure \ref{fig:sim_spec_tune} (Appendix \ref{sec:app_multicomp}).  From this, it is clear that we are, to some extent, able to distinguish the mixing ratios of different ice components.  High H$_2$O:CO$_2$ ratios (10:3) provide poorer agreement with the observed ice band profiles.  Moreover, the observed CO band profile is incompatible with a pure-CO component.  We cannot, however, clearly distinguish between lower mixing ratios for polar mixtures (10:1 vs.~10:2 H$_2$O:CO$_2$ and H$_2$O:CO), nor precisely constrain the CO$_2$:CO mixing ratio.  We subsequently adopt the model presented in Figure \ref{fig:sim_spec} (bottom) as our preferred ice composition due to the good agreement with the observed band profiles, but emphasize that somewhat different mixing ratios are still compatible with the observations. 

The main differences between our preferred model (Figure \ref{fig:sim_spec} bottom) and the observations are: the excess absorption features on the H$_2$O band (attributable to trace solid species); the CO$_2$ wings (due to scattering); and the blue CO$_2$ shoulder around 4.23 $\mu$m.  We are unable to reproduce this shoulder with any combination of the mixed and pure ices considered so far, and address its possible origin in more detail in Appendix \ref{subsec:app_co2wing}.

It is important to recognize that, while the line cores should be less impacted by scattering effects than the line wings, they may still be somewhat affected.  It therefore remains possible that different assumptions about e.g.~the grain distribution would bring model S3 (separate polar/apolar) into better agreement with the observations.  Given our current model, however, a scenario including a polar ice component (such as M1) can better describe the ice band core profiles compared to apolar-only CO and CO$_2$ ice (S3).

\subsection{The role of CH$_3$OH}
\label{subsec:res_ch3oh}

\begin{figure*}
\centering
    \includegraphics[width=\linewidth]{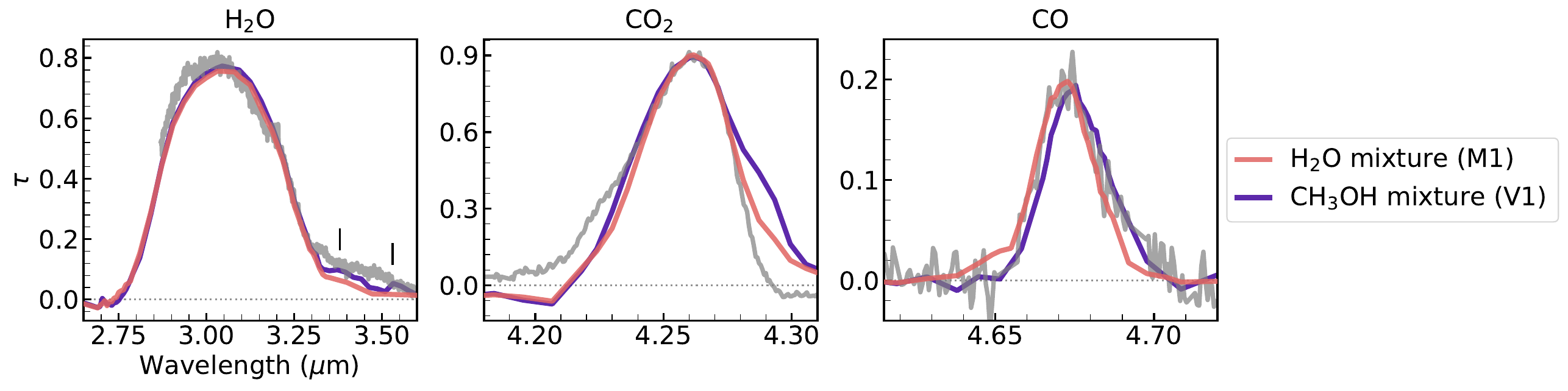}
    \caption{Comparison of band profiles for different polar mixtures of CO$_2$ and CO.  The `H$_2$O mixture' spectrum is the preferred multi-component model shown in Figure \ref{fig:sim_spec} bottom.  In the `CH$_3$OH mixture' spectrum, the H$_2$O:CO$_2$ and H$_2$O:CO components are replaced by CH$_3$OH:CO$_2$ and CH$_3$OH:CO.  In the left panel, the positions of CH$_3$OH absorption features are indicated with black ticks.  Ice abundances for each model are listed in Table \ref{tab:abundances} with the model names in parentheses.}  
    \label{fig:ch3oh}
\end{figure*}

So far we have only attempted to fit the observed ice bands using combinations of the major ice species H$_2$O, CO$_2$, and CO.  Here, we briefly consider whether we can discriminate the presence of CH$_3$OH based on its influence on the CO$_2$ and CO bands.  In Figure \ref{fig:ch3oh}, we show the simulated ice bands starting with the preferred multi-component model from Section \ref{subsec:res_finetune}, but assuming that the polar CO and CO$_2$ components are mixed with CH$_3$OH instead of H$_2$O.  We used laboratory spectra with a $\sim$1:1 mixing ratio of CH$_3$OH with CO$_2$ and CO (experiments 11 and 12).  For this test, we distributed the CO$_2$:CH$_3$OH mixture along with CO$_2$ (zones 2 and 3) and the CO:CH$_3$OH mixture along with CO (zone 3).  While CH$_3$OH has a sublimation temperature close to H$_2$O and can trap more volatile molecules \citep{Burke2015}, we do not expect entrapment to be efficient for low CH$_3$OH mixing fractions \citep{Simon2019}.  

The presence of CH$_3$OH results in significant broadening of the CO$_2$ band and is clearly incompatible with the observed JWST spectrum.  Given this, it is unlikely that the CO$_2$ ice in HH 48 NE contains an appreciable amount of CH$_3$OH.  On the other hand, replacing the CO:H$_2$O component with a CO:CH$_3$OH component produces a band profile that is more compatible with the observed spectrum.  Compared to the CO:H$_2$O mixture, the band profile for the CH$_3$OH mixture agrees better with the observed spectrum in the wings though somewhat worse in the core.  For CO, we therefore consider CH$_3$OH to be a comparably likely polar component as H$_2$O, or possibly a combination of CO:CH$_3$OH and CO:H$_2$O is present.  Indeed, ice mixtures of CO with CH$_3$OH instead of H$_2$O were previously proposed as a solution to the non-detection of the blue CO shoulder in interstellar ice observations \citep{Cuppen2011}.  

We tested mixtures with relatively high CH$_3$OH fractions (1:1 relative to CO$_2$ and CO), and it is important to check whether this over-produces CH$_3$OH absorption compared to the observed spectrum.  In Figure \ref{fig:ch3oh}, the CH$_3$OH CH$_3$ stretching modes around 3.4 and 3.53 $\mu$m are visible in the simulated spectrum, but at levels below the red shoulder seen on the H$_2$O band.  It is therefore plausible for the ices to contain these CH$_3$OH abundances without being in conflict with the observations.  

\begin{figure*}
\centering
    \includegraphics[width=\linewidth]{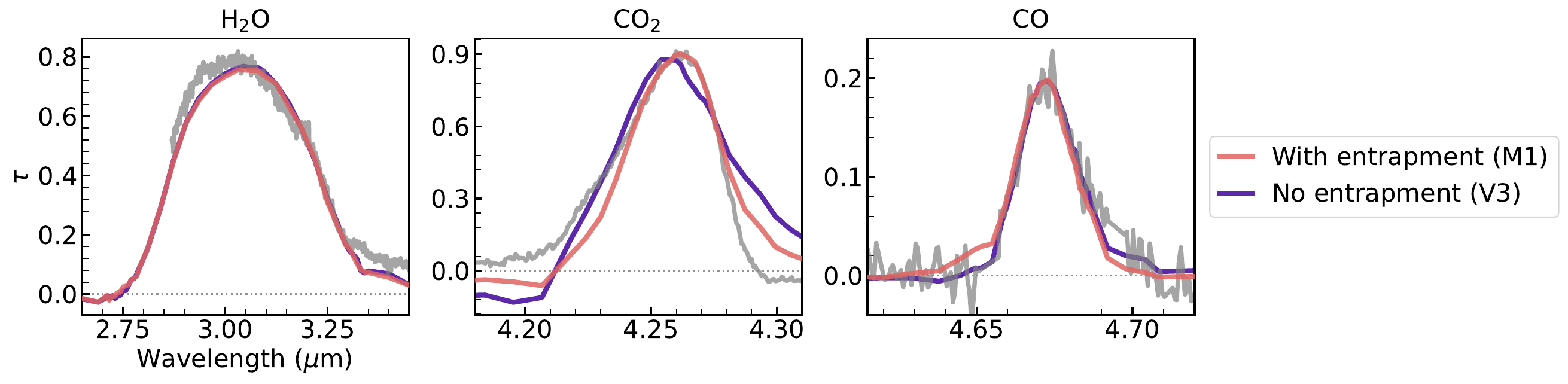}
    \caption{Comparison of the preferred multi-component ice model from Section \ref{subsec:res_finetune} with a model in which CO and CO$_2$ are not trapped interior to their sublimation fronts.  The `no entrapment' model contains CO$_2$:CO and pure CO in zone 3, along with pure CO$_2$ in zones 2-3 and pure H$_2$O in zones 1-3.  Ice abundances for each model are listed in Table \ref{tab:abundances} with the model names in parentheses.}  
    \label{fig:co_notrap}
\end{figure*}

\subsection{The role of volatile trapping}
\label{subsec:res_trapping}
It is clear from our modeling that mixed ices are needed to reproduce the HH 48 NE band profiles.  So far, we have assumed that mixtures imply trapping, i.e.~that in H$_2$O- and CO$_2$-dominated mixtures, small amounts of CO will be trapped until the matrix sublimates.  Given the implications of volatile entrapment (Section \ref{subsec:disc_c/o}), it is important to determine if we can in fact distinguish this scenario from a model with the same ice mixtures present, but where CO-containing ices are confined to the CO freeze-out zone.  

Figure \ref{fig:co_notrap} shows the resulting `no entrapment' model compared to our preferred model from Section \ref{subsec:res_finetune}.  The H$_2$O and CO bands are reproduced well in the no-entrapment model.  However, the CO$_2$ band is significantly broader with more pronounced wings, producing a poorer fit to the observations than the entrapment model.  The broad CO$_2$ component arises from the high abundance of CO$_2$:CO ice in zone 3, which is needed to reproduce both the intensity and shape of the CO band.  We suspect that the increased importance of large grains within the CO freeze-out zone significantly alters the CO$_2$ band profile, which is particularly sensitive to grain size effects \citep{Dartois2022}.  Therefore, reproducing the CO band profile in an entrapment-free model requires high CO$_2$ abundances in the CO freeze-out zone, which produces a broad CO$_2$ band profile incompatible with the observations.  This provides compelling observational evidence for CO trapping in CO$_2$ ice.  Coverage of the 2.75 $\mu$m dangling OH band region at high SNR would help to assess the role of CO and CO$_2$ trapping in H$_2$O ice.

\subsection{CO$_2$ saturation and the $^{12}$CO$_2$/$^{13}$CO$_2$ ratio}
\label{subsec:res_13co2}
$^{13}$CO$_2$ ice was also detected towards HH 48 NE, with an integrated absorption ratio 1/14 that of $^{12}$CO$_2$ \citep{Sturm2023c}.  Being a minor component of the ice, $^{13}$CO$_2$ is less affected by grain shape effects compared to the main isotopologue \citep{Boogert2000}, and can help to validate our retrievals from $^{12}$CO$_2$.  We simulated the $^{13}$CO$_2$ band by scaling down the CO$_2$ ice abundance.  For simplicity we ray-traced using opacities for a pure $^{12}$CO$_2$ ice, which we then shifted to the wavelength position of $^{13}$CO$_2$.  This approach is reasonable given that $^{12}$CO$_2$ and $^{13}$CO$_2$ have similar band strengths, and the radiation propagation should not differ significantly between the positions of the two bands (4.28 and 4.4$\mu$m).

\begin{figure}
\centering
    \includegraphics[width=\linewidth]{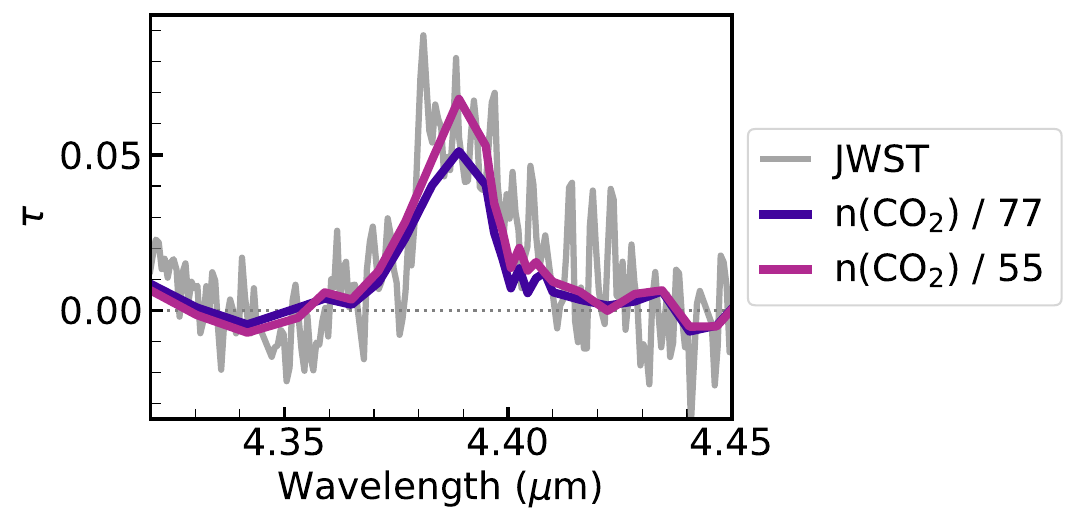}
    \caption{The $^{13}$CO$_2$ ice band observed with JWST (grey) and simulated (colors).  The $^{13}$CO$_2$ abundances in the simulations are determined by scaling down the CO$_2$ abundance from the pure ice model (S1) by the factors indicated.}  
    \label{fig:13CO2}
\end{figure}
The resulting simulated $^{13}$CO$_2$ spectra are shown in Figure \ref{fig:13CO2}, with the observed spectrum shown for comparison.  We tested both an ISM-like $^{12}$C/$^{13}$C ratio of 77 \citep{Wilson1999}, as well as a lower value of 55 compatible with the low end of the range of $^{12}$CO$_2$/$^{13}$CO$_2$ ratios seen towards interstellar ices \citep{Boogert2000}.  The ratio of 55 closely reproduces the optical depth of the observed band.  While this should not be taken as a fit, it demonstrates that extreme carbon fractionation is not necessary to explain the detection of the $^{13}$CO$_2$ ice band in HH 48 NE.  Instead, this confirms that the $^{12}$CO$_2$ band is quite optically thick \citep{Sturm2023b, Sturm2023c}.  Higher signal-to-noise coverage of the $^{13}$CO$_2$ feature would help to more firmly extract a $^{12}$CO$_2$/$^{13}$CO$_2$ ratio. 

\section{Discussion}
\label{sec:discussion}
We have performed a detailed analysis of the H$_2$O, CO$_2$, and CO ice band profiles observed towards the edge-on disk HH 48 NE with JWST, retrieving the first constraints on the mixing of these major ice species in a protoplanetary disk.  Here we discuss the implications for the icy environment of planetesimal and planet formation.  It is important to emphasize that the absolute ice abundances determined from this modeling are sensitive to the adopted disk physical structure and ice distribution, for which there remain some uncertainties for HH 48 NE \citep{Sturm2024}.  Moreover, due to the partitioning of ice onto different grain populations with different vertical distributions, the local ice/H abundances and ice:rock ratios can vary considerably across the disk model, making it challenging to interpret the retrieved ice abundances in an absolute sense.  We therefore focus on comparisons of the \textit{ratios} of different ice species and mixture components, which should be relatively insensitive to these considerations.

\subsection{Ice compositions}
\label{subsec:disc_abunds}

\begin{figure}
\centering
    \includegraphics[width=\linewidth]{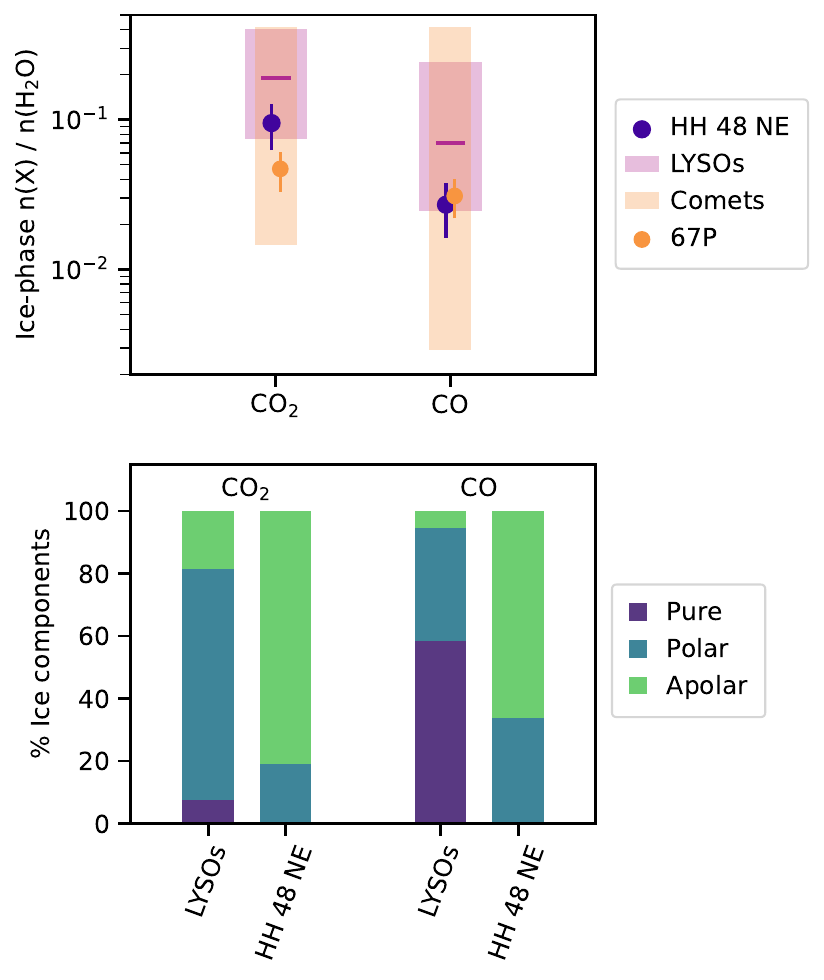}
    \caption{Comparison of ice compositions across different evolutionary stages.  Top: ratios of H$_2$O, CO$_2$, and CO with respect to H$_2$O, shown for HH 48 NE as well as low-mass young stellar objects (LYSOs) and solar system comets.  For LYSOs \citep{Boogert2015}, the median is shown with a pink line and the 25th--75th percentile range with a pink shaded bar.  The ranges seen in solar system comets \citep{Mumma2011} are shown with an orange shaded bar, and those for 67P specifically  \citep{Rubin2019} with an orange circle.  Bottom: proportion of pure, polar, and apolar components of CO$_2$ and CO measured towards LYSOs (median across the sample in \citealt{Oberg2011}) and HH 48 NE.}  
    \label{fig:ice_abunds}
\end{figure}

Our measurements towards HH 48 NE provide the first opportunity to compare the evolution of ice compositions from protostars to disks to planetesimals.  Figure \ref{fig:ice_abunds} (top) shows the CO$_2$/H$_2$O and CO/H$_2$O ratios in HH 48 NE as well as a sample of low-mass young stellar objects (LYSOs; \citealt{Boogert2015}) and solar system comets \citep{Mumma2011}.  We also include specifically the abundances derived from ROSINA measurements of comet 67P/Churymov-Gerasimenko (hereafter 67P; \citealt{Rubin2019}).  Uncertainties on the abundance ratios in HH 48 NE are estimated as described in Appendix \ref{sec:app_abund}.  Note that the ratios for HH 48 NE are only strictly valid in the disk zones 2 and 3 where all volatiles in the model are frozen out.  

The median LYSO CO$_2$/H$_2$O and CO/H$_2$O ratios are somewhat higher than those inferred towards HH 48 NE.  This is suggestive of a loss of the relatively volatile species (CO$_2$ and CO) from the ice during the initial stages of disk formation and grain growth.  The CO$_2$ ice abundance appears to decrease by another factor of $\sim$2 going from HH 48 NE to comet 67P, while the CO ice abundance is similar in both cases.  There is a large spread in CO$_2$/H$_2$O and CO/H$_2$O ratios measured towards other comets, making it impossible to draw conclusions about volatile loss after disk formation.  Nonetheless, the lower CO$_2$/H$_2$O and CO/H$_2$O ratios in HH 48 NE compared to typical low-mass protostars is suggestive that some hypervolatile loss occurs prior to planetesimal formation and evolution.  Better demographics for the ice abundances in disks and comets are needed to statistically constrain the relationship between these populations.

\subsection{Ice mixing status}
\label{subsec:disc_comp}

CO$_2$ and CO ices in dense cloud and protostellar environments are commonly found to consist of a mixture of pure, polar, and apolar components \citep[e.g.][]{Tielens1991,Chiar1995,Chiar1998, Pontoppidan2003,Pontoppidan2008}.  Determining whether this structure is preserved in disk ices can help to shed light on the degree of processing experienced by ice mantles during incorporation into the disk.  Figure \ref{fig:ice_abunds} (bottom) shows the median proportion of pure, polar, and apolar components of CO$_2$ and CO measured towards low-mass young stellar objects \citep{Oberg2011}, along with the ratios retrieved for HH 48 NE.  

In HH 48 NE, we infer that CO$_2$ and CO are mostly present in an apolar mixture, with no pure CO or CO$_2$ component.  Our model also favors small components of polar CO$_2$ and CO, though the exact contribution could be influenced by the way in which scattering is treated (Section \ref{subsec:res_finetune}).  From this, it is clear that the HH 48 NE ice composition is not directly inherited from an earlier evolutionary stage.  Notably, the apolar mixture represents much more of the ice budget in HH 48 NE compared to low-mass protostars.  A possible explanation is thermal processing leading to segregation or distillation of CO and CO$_2$ from the polar layer \citep{Ehrenfreund1998, Oberg2009, Brunken2024}.  Chemical processing of the pure-CO component seen in protostars into a CO$_2$:CO mixture \citep{Ioppolo2011} could also contribute to the growth of the apolar layer.  It is also important to recognize that the lack of pure CO inferred for HH 48 NE may reflect an observational bias: photons escaping the disk likely trace a somewhat elevated vertical layer of the disk \citep{Sturm2024}, and therefore the observed spectrum may not be sensitive to a pure-CO reservoir below the CO snow surface.  

With the current observational constraints, the exact nature of the disk ice polar component is not entirely clear.  Mixtures with either H$_2$O or CH$_3$OH both provide a reasonable fit to the CO band profile.  There are plausible astrochemical pathways for CO to be mixed with either H$_2$O or CH$_3$OH: either via CO accretion during the formation of H$_2$O-dominated ice in conditions with high H/CO ratios; or via formation of CH$_3$OH from CO hydrogenation following catastrophic CO freeze-out in conditions with low H/CO ratios \citep[e.g.][]{Tielens1991, Cuppen2009}.  On the other hand, the CO$_2$ band profile clearly excludes mixing of CO$_2$ with CH$_3$OH (Section \ref{subsec:res_ch3oh}).  Experiments support that CO$_2$ can form either within CO-rich ices \citep{Ioppolo2011}, or in the absence of CO via radiation processing of carbon grains or large organics \citep[e.g.][]{Mennella2004,Mate2024}.  It is therefore plausible either that CO and CO$_2$ occupy the same polar phase (with H$_2$O), or different polar phases (CO with CH$_3$OH and CO$_2$ with H$_2$O).  While CH$_3$OH was not detected at 3.54 $\mu$m \citep{Sturm2023c}, future analysis of the mid-IR spectrum may provide improved constraints on the CH$_3$OH ice budget and in turn the feasibility of CO:CH$_3$OH mixtures.  Meanwhile, coverage of the 2.75 $\mu$m dangling OH feature is needed to quantify the extent to which volatiles like CO and CO$_2$ are present in the water matrix.

\subsection{C/O ratio of icy solids}
\label{subsec:disc_c/o}

\begin{figure}
\centering
    \includegraphics[width=\linewidth]{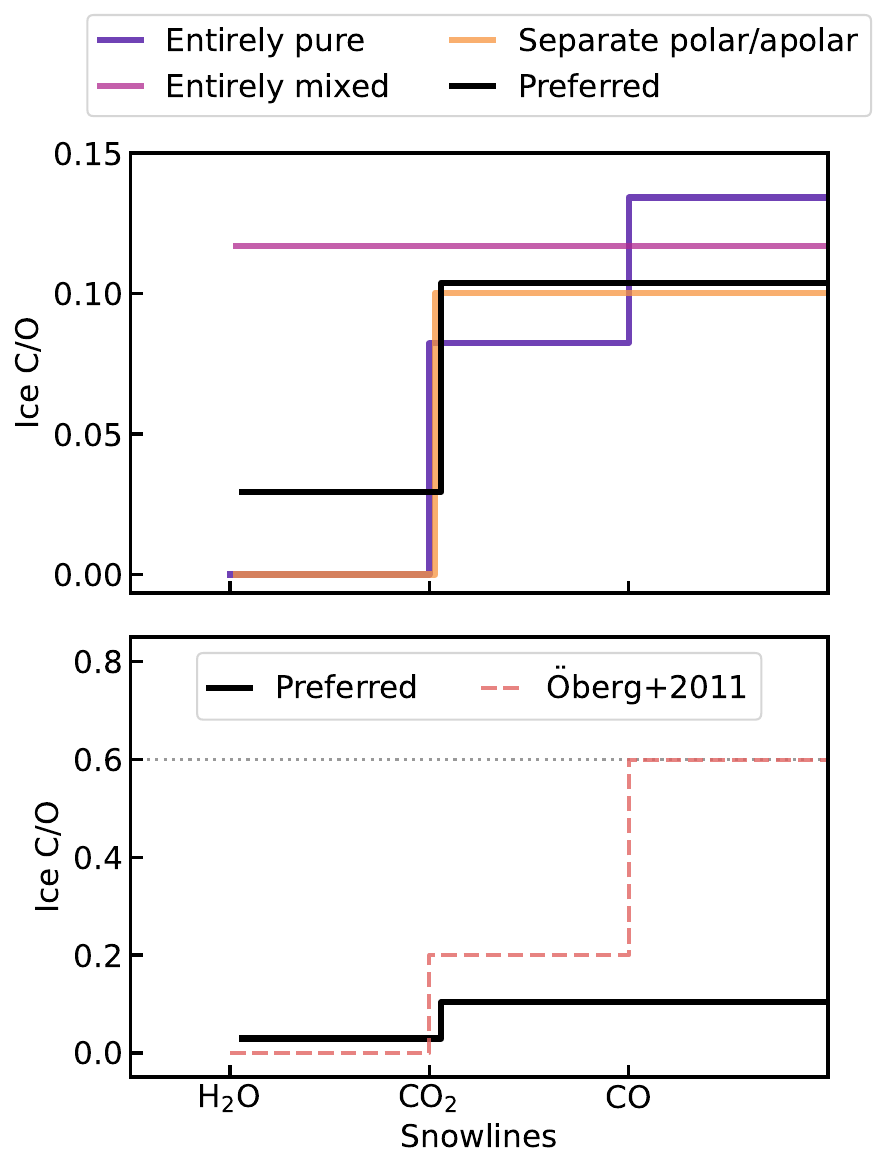}
    \caption{Top: ice-phase C/O ratios retrieved from the single-component and preferred multi-component ice models.  Bottom: comparison of HH 48 NE ice C/O ratios to the ice C/O ratios in the model of \citet{Oberg2011b}.}  
    \label{fig:co_ratios}
\end{figure}

H$_2$O, CO$_2$, and CO represent the major carbon and oxygen-bearing species in dense star-forming regions, and together will largely determine the elemental C/O ratio of the ice.  We can directly extract the ice-phase C/O ratios from our models, with different disk zones corresponding to the regions defined by the H$_2$O, CO$_2$, and CO sublimation fronts.  Figure \ref{fig:co_ratios} top shows the ice-phase C/O ratios for the single-component models (Section \ref{subsec:res_endmemb}) as well as the preferred multi-component model (Section \ref{subsec:res_finetune}).  The bottom panel shows the ice C/O ratios from our preferred model compared to the canonical values in \citet{Oberg2011b}.  Because of uncertainties in the ice:rock ratios, for both our results and for the \citet{Oberg2011b} model, we show only ice-phase C/O ratios (i.e.~we do not account for C and O contained in refractory grains).  An important caveat is that these observations likely do not trace the disk midplane, and therefore may underestimate in particular the abundance of pure CO ice.  Nonetheless, we can draw several interesting conclusions even with this uncertainty in mind.

First, it is important to note that the radial variation of C/O depends on the structure of the ice in addition to the ice abundances.  While the single-component models are overly simplistic, they demonstrate that more ice mixing and volatile entrapment will decrease the importance of the snowlines, leading to smaller radial variations in the ice C/O ratio.  By extension, this will also reduce radial variations in the gas-phase C/O ratio, since entrapped volatiles will not contribute to the C and O budgets in the gas phase.  This underscores the importance of constraining ice mixing status for accurately predicting radial variations in the gas and ice C/O ratios.  Doing so is critical to our ability to correctly interpret and predict the compositions of exoplanet atmospheres, hydrospheres, and cryospheres \citep[e.g.][]{Marboeuf2008, Oberg2011b, Johnson2012, Marounina2020}.

Our preferred ice model contains CO and CO$_2$ trapped in water.  This results in a non-zero ice-phase C/O ratio between the H$_2$O and CO$_2$ snowlines, with solids containing more volatile carbon in this region than would be expected for a pure-ice scenario.  Additionally, the ice C/O ratio we retrieve exterior to the CO snowline (0.1) is much lower than the range of $\sim$0.35--0.6 predicted from protostellar ice abundances \citep[e.g.][]{Oberg2011,Oberg2011b}.  One possibility is that pure CO ice in the midplane could explain this difference, which would necessitate an additional midplane CO ice abundance around 10$^{-4}$ n$_\mathrm{H}$, i.e.~a significant quantity of pure CO.  Alternatively, speciation of O into H$_2$O rather than CO$_2$ and CO within the disk could suppress the ice-phase C/O while boosting the gas-phase C/O.  Indeed, coupled photochemistry and dynamics within the disk gas are expected to sequester H$_2$O in ices while increasing the gas-phase carbon content \citep{Bergin2016, vanClepper2022}.  It is reasonable to expect that a combination of these factors may be at play, i.e.~additional carbon-bearing ices are hidden in the midplane \textit{and} ice-phase H$_2$O is enhanced over CO and CO$_2$ due to photochemical-dynamical evolution.

\subsection{Model limitations \& outlook}
\label{subsec:disc_caveats}

The framework presented in this work represents an important step forward in simultaneously accounting for ice abundances, distributions, mixing statuses, and radiative transfer effects in modeling ice spectra of edge-on disks.  Given the complexity of this problem, we still had to make some assumptions and approximations, most notably fixing disk physical attributes like the grain size distribution and the vertical ice extent.  Firm constraints on the absolute abundances of disk ices will require these to be considered in tandem with the ice compositions.  Nonetheless, with this model we are able to provide a first look at the ice mixtures present within the HH 48 NE disk and the importance of volatile entrapment in shaping the disk volatile budget.

In future work, it should be possible to obtain independent constraints on the sizes of ice-coated dust grains using the CO$_2$ line wings, which are highly sensitive to the grain size distribution \citep{Dartois2022}.  Additionally, the ice vertical distribution can be probed by considering spatial variations in the observed ice spectra \citep{Sturm2024}.  More sophisticated treatments of the ice opacities, including e.g.~varying ice temperatures across the disk, will also enable more robust comparisons with the observed band profiles.  Improvements in the observations will also be important: higher-SNR data is needed to perform band profile analysis in a spatially resolved way, particularly for the CO band.  And, expanded wavelength coverage across the near- to mid-IR will provide new leverage for constraining both disk physical structures and ice compositions.  Notably, it is clear from our modeling that H$_2$O mixing cannot be inferred without the dangling OH feature at 2.75 $\mu$m.  Also, the 15.2 $\mu$m CO$_2$ bending mode is a more definitive indicator of CO$_2$ ice mixing compared to the 4.25~$\mu$m stretching mode \citep{Pontoppidan2008}.  

\section{Conclusions}
\label{sec:concl}
The modeling framework introduced here has provided the first constraints on the ice composition and mixing within a protoplanetary disk.  Our key findings are as follows:
\begin{itemize}[noitemsep,leftmargin=0.2in]
    \item Most CO$_2$ and CO ices are present in an apolar mixture.  There is compelling spectroscopic evidence that CO is not only mixed with, but entrapped within, CO$_2$ ice.
    \item Our models favor that there is also a small component of polar CO$_2$ and CO.  From the observed band profiles, CO can be mixed with either H$_2$O or CH$_3$OH, whereas a CO$_2$ mixture with CH$_3$OH is disfavored.
    \item The mixing status of H$_2$O cannot be assessed without coverage of the 2.75 $\mu$m dangling OH band.  
    \item The retrieved CO/H$_2$O and CO$_2$/H$_2$O ratios for HH 48 NE appear lower than the median values observed towards low-mass young stellar objects, hinting that some hypervolatile loss may occur prior to the formation of planetesimals.
    \item Compared to typical low-mass protostars, the HH 48 NE disk has a higher proportion of apolar CO$_2$ and CO and a lower proportion of polar CO$_2$ and CO.  This implies thermal and/or chemical processing of the ice during disk formation or evolution.
    \item We do not detect any pure CO ice component, though this may be an observational bias if the edge-on disk spectra do not trace elevations below the CO snow surface.
    \item An ISM-like $^{12}$CO$_2$/$^{13}$CO$_2$ abundance ratio can reproduce the observed optical depth ratio due to saturation of the $^{12}$CO$_2$ band.
    \item Different scenarios for ice mixing and entrapment result in different radial variations in the ice-phase C/O ratio, underscoring the importance of ice microphysics to the partitioning of volatiles within a disk.
    \item We infer a low ice-phase C/O ratio of $\sim$0.1 throughout the disk.  This could reflect a combination of (i) a hidden midplane reservoir rich in pure CO ice, and (ii) fractionation into an O-rich ice and C-rich gas due to photochemical-dynamical evolution during the disk stage.
\end{itemize}

JWST has opened a new window on icy volatiles in protoplanetary disks.  These early findings reveal that protostellar ice compositions are not inherited completely intact to the disk stage.  Moreover, ice mixing and entrapment must be accounted for in efforts to understand the distribution of icy volatiles across the disk.  This is of critical importance to predicting and interpreting the compositions of exoplanet atmospheres, hydrospheres, and cryospheres.  JWST observations of $>$10 additional edge-on disks are ongoing or scheduled (Programs GTO 1282, GO 1621, GO 1751, GO 5299), which will provide a much-needed demographic picture of the icy landscape in which planets assemble

\section*{acknowledgements}

We are grateful to the anonymous referee for their valuable feedback on this manuscript.  
M.N.D. acknowledges the Holcim Foundation Stipend.
Support for C.J.L. was provided by NASA through the NASA Hubble Fellowship grant No. HST-HF2-51535.001-A awarded by the Space Telescope Science Institute, which is operated by the Association of Universities for Research in Astronomy, Inc., for NASA, under contract NAS5-26555.
 We acknowledge AJ Galaviz III (Southwest Research Institute) for his Graphics, enriching our space programs research with visual clarity and creativity.
S.I. acknowledges support from the Danish National Research Foundation through the Centre of Excellence 'InterCat' (grant agreement No. DNRF150).
Part of this research was carried out at the Jet Propulsion Laboratory, California Institute of Technology, under a contract with the National Aeronautics and Space Administration (80NM0018D0004).
J.A.N. and E.D. acknowledge support from French Programme National `Physique et Chimie du Milieu Interstellaire' (PCMI) of the CNRS/INSU with the INC/INP, co-funded by the CEA and the CNES.
D.H. is supported by a Center for Informatics and Computation in Astronomy (CICA) grant and grant number 110J0353I9 from the Ministry of Education of Taiwan. D.H. also acknowledges support from the National Science and Technology Council of Taiwan through grant number 111B3005191.
%\end{acknowledgements}

\software{
{\fontfamily{qcr}\selectfont RADMC-3D} \citep{Dullemond2012},
{\fontfamily{qcr}\selectfont optool} \citep{Dominik2021},
{\fontfamily{qcr}\selectfont icenk} \citep{Gerakines2020},
{\fontfamily{qcr}\selectfont Matplotlib} \citep{Hunter2007},
{\fontfamily{qcr}\selectfont NumPy} \citep{VanDerWalt2011},
{\fontfamily{qcr}\selectfont Scipy} \citep{SciPy2020},
}

\bibliography{references}

\appendix
\FloatBarrier
\section{Ice spectrum interpolation}
\label{sec:app_interp}

\begin{figure}[h]
\centering
    \includegraphics[width=\linewidth]{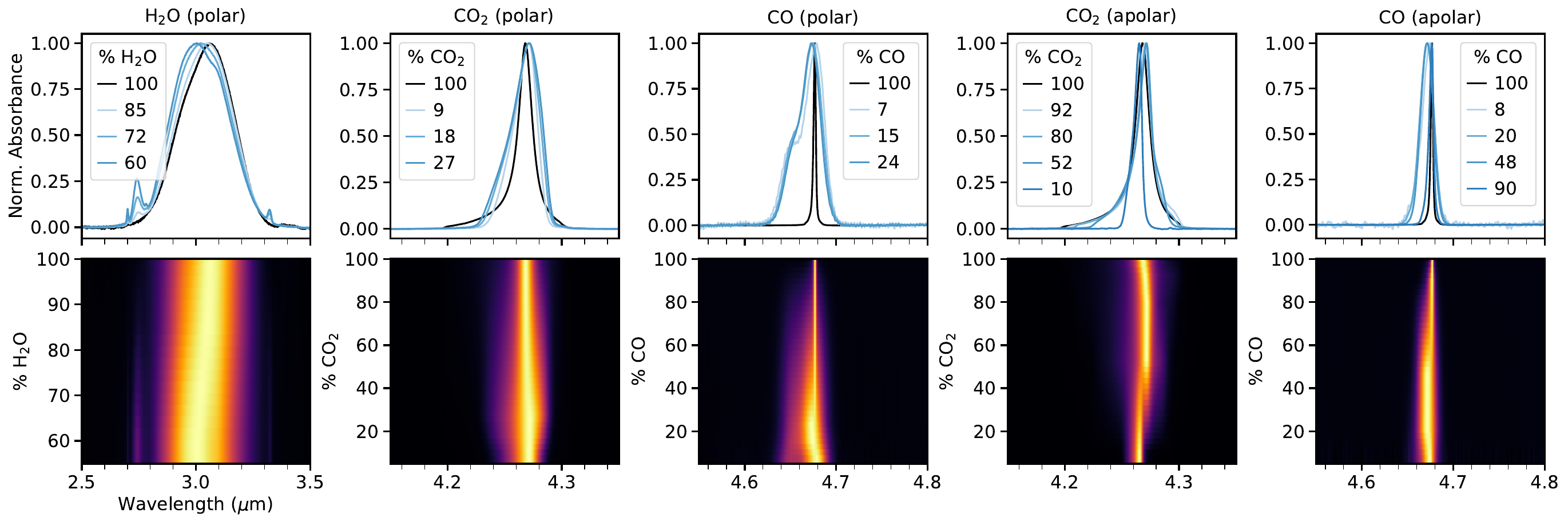}
    \caption{Illustration of ice spectrum interpolation.  The relative absorbance of an ice band can be expressed as a function of wavelength and mixing ratio with other ice components.  Top: experimental ice profiles for different mixing ratios.  Bottom: 2D interpolation map of normalized band profiles verses mixing ratio.}  
    \label{fig:interp}
\end{figure}

In order to fit the observed ice bands, we require reference spectra with a range of H$_2$O:CO$_2$:CO abundances and mixing ratios.  Since it is impractical to obtain dozens of experimental reference spectra with slightly varying mixing ratios, we performed an interpolation using the ice compositions described in Table \ref{tab:exps}.  For each of the H$_2$O, CO$_2$, and CO ice bands, the normalized band profile can be expressed as a function of (i) wavelength and (ii) mixing ratio with other ice species.  The spectral profiles for each mixing ratio are shown in Figure \ref{fig:interp} (top).  2D interpolations of relative absorbance as a function of wavelength and mixing ratio are shown in Figure \ref{fig:interp} (bottom).  Polar and apolar mixtures are interpolated separately.  The mixing ratio of H$_2$O is determined from N(H$_2$O)/N(H$_2$O+CO$_2$+CO).  The CO$_2$ and CO ratios in polar ices are determined from N(CO$_2$)/N(H$_2$O+CO$_2$) and N(CO)/N(H$_2$O+CO), respectively.  While this neglects the mutual influence of CO$_2$ and CO on each other within a H$_2$O-dominated mixture, we expect this to be a minor effect, especially for the low mixing ratios used for our preferred models (10:1 H$_2$O:CO$_2$ and 10:1 H$_2$O:CO).  Treating the polar CO$_2$ and CO components separately also allows us to independently tune their abundances, which is necessary for reproducing the observed spectra.  

\begin{figure}
\centering
    \includegraphics[width=\linewidth]{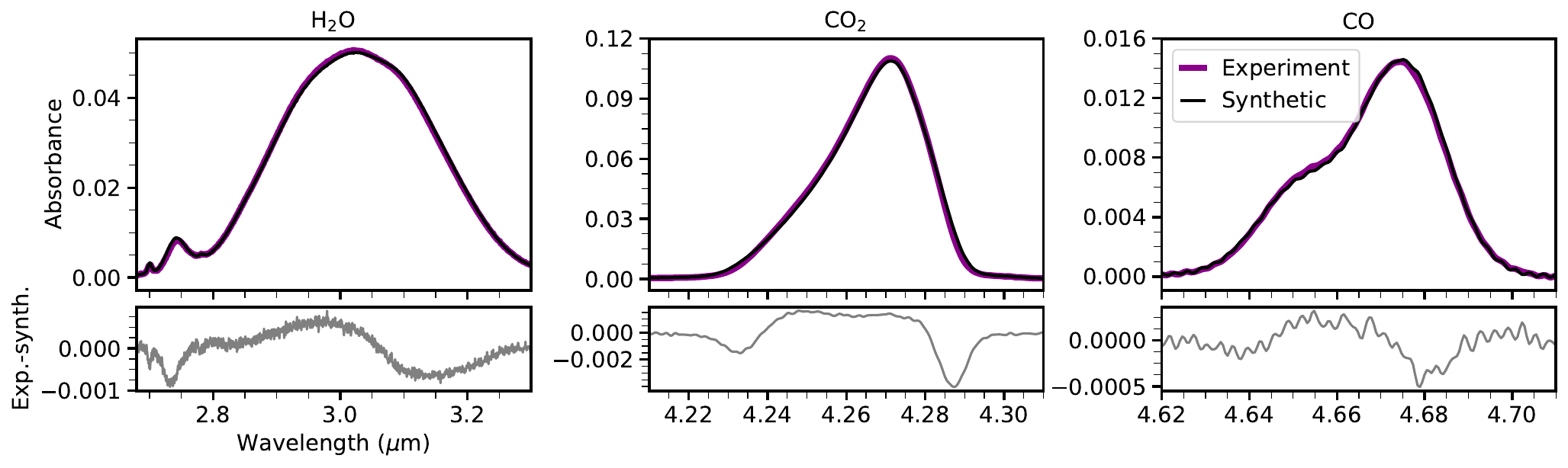}
    \caption{Test of ice spectrum interpolation scheme.  Top: ice absorption bands for a $\sim$10:2:2 H$_2$O:CO$_2$:CO mixture.  The experimental spectrum (Exp.~5) is shown in pink.  The synthetic spectrum (black) is made by interpolating between the 10:1:1 and 10:3:3 spectra.  Bottom: Differences between the synthetic and experimental spectrum.}  
    \label{fig:test_interp}
\end{figure}

With this approach, synthetic spectra can be generated for arbitrary mixing ratios of H$_2$O:CO$_2$:CO using the interpolated band profiles.  The absolute absorbance is determined by scaling the integrated band area to match the desired column density for each molecule.  We have verified this procedure by interpolating only the polar 10:1:1 and 10:3:3 spectra and then attempting to reproduce the 10:2:2 spectrum.  Even for this wide gap in measured mixing ratios, the synthetic spectrum provides excellent agreement with the measured spectrum ($\lesssim$5\% deviations; Figure \ref{fig:test_interp}).  

\FloatBarrier
\section{Radiative transfer spectra}
\label{sec:app_RT}

\subsection{Comparison of multi-component ice compositions}
\label{sec:app_multicomp}
\begin{figure}
\centering
    \includegraphics[width=\linewidth]{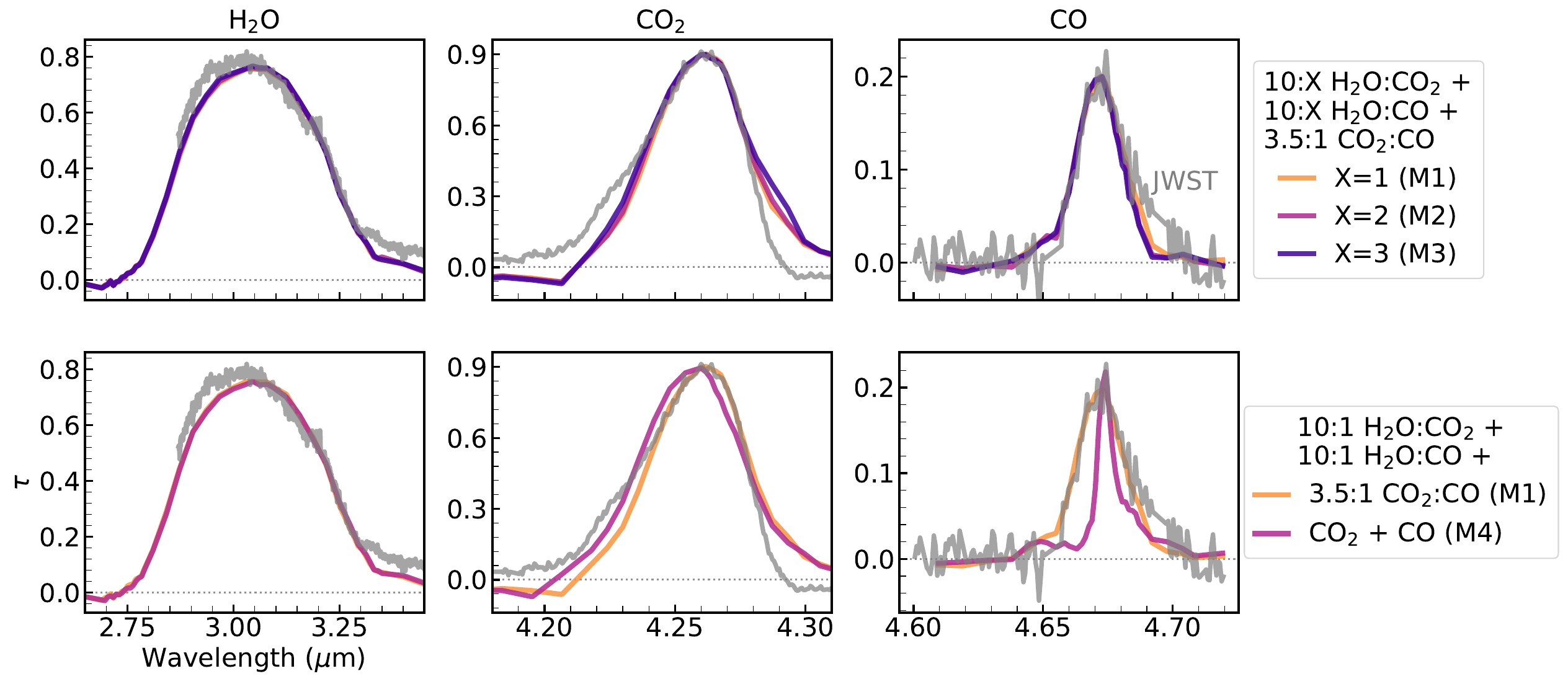}
    \caption{Simulated HH 48 NE ice band profiles (colors) compared to the spectrum observed with JWST (grey).  Top: varying mixing ratios within the polar ice components (H$_2$O:CO$_2$ and H$_2$O:CO).  Bottom: varying mixing ratios within the apolar ice component (CO$_2$ and CO).  All spectra include an additional H$_2$O-only component.  Ice abundances for each model are listed in Table \ref{tab:abundances} with the model names in parentheses.  Model M1 corresponds to the preferred multi-component ice composition.}  
    \label{fig:sim_spec_tune}
\end{figure}

Figure \ref{fig:sim_spec_tune} shows a comparison of different multi-component ice mixtures, introduced in Section \ref{subsec:res_finetune}.  All models include H$_2$O, polar CO$_2$ and CO in H$_2$O-dominated mixtures, and an apolar component of CO and CO$_2$ either mixed together or separate.  This combination of components can better reproduce the observed band profiles than pure, polar, or apolar ices individually.  We test several mixing ratios for both the polar and apolar mixtures.  In all cases, we aim to reproduce the peak optical depths of the H$_2$O, CO$_2$, and CO ice bands.  The top row of Figure \ref{fig:sim_spec_tune} shows H$_2$O:CO$_2$ and H$_2$O:CO mixing ratios ranging from 10:1 to 10:3.  While we cannot distinguish the low CO$_2$ fractions (10:1 vs.~10:2), the red CO$_2$ wing becomes more pronounced for a higher CO$_2$ fraction of 10:3, providing worse agreement with the observed spectrum.  We therefore favor a lower mixing ratio in our preferred model.  We adopt a ratio of 10:1 for subsequent modeling.

The bottom row of Figure \ref{fig:sim_spec_tune} shows different scenarios for the apolar CO$_2$+CO: either a CO$_2$-dominated mixture or non-interacting CO$_2$ and CO components.  Note that entrapment would not be possible for a CO-dominated mixture of CO$_2$:CO, and this scenario is therefore ruled out based on the same arguments as in Section \ref{subsec:res_trapping}.  The CO$_2$:CO mixture (M1) provides a better match to the core of the observed CO$_2$ band compared to the pure-CO$_2$ model (M4), which produces too much absorption on the blue side of the feature.  The observed CO band profile is incompatible with a pure-CO component.  We therefore adopt the CO$_2$:CO mixture as our preferred form of apolar CO$_2$ and CO, and rule out that there is an appreciable contribution of pure CO to the observed spectrum.  A $\sim$3.5:1 mixture is needed to reproduce the optical depths of both the CO$_2$ and CO bands.

\FloatBarrier

\subsection{The CO$_2$ blue shoulder}
\label{subsec:app_co2wing}
\begin{figure}
\centering
\includegraphics[width=0.5\linewidth]{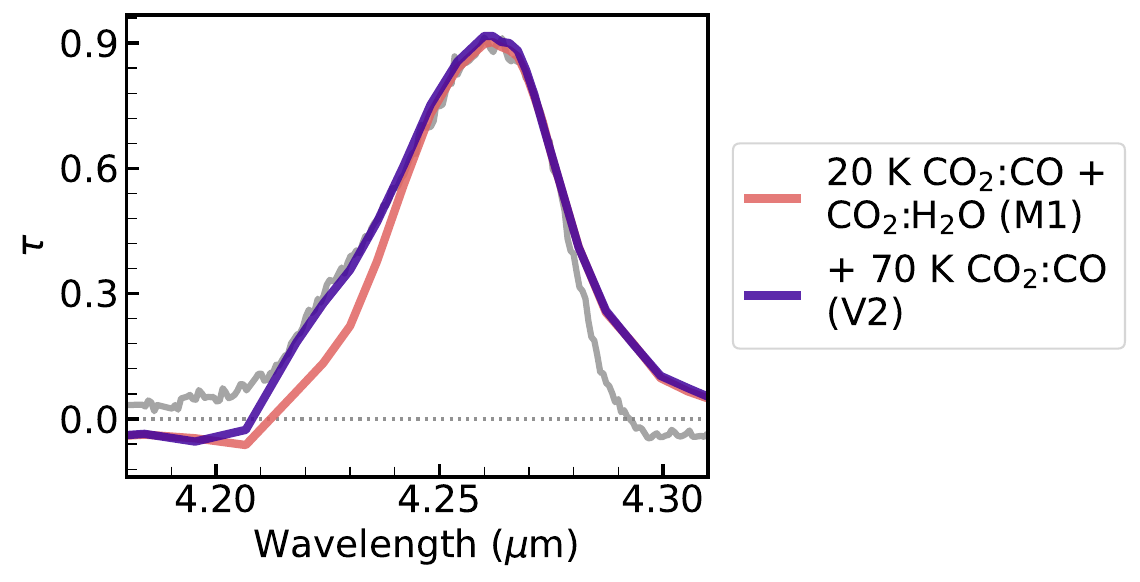}
    \caption{The CO$_2$ ice band observed with JWST (grey) compared to the preferred multi-component ice mixture from Section \ref{subsec:res_finetune} (orange), and including an additional component of warmed CO$_2$:CO ice (purple).}  
    \label{fig:CO2_shoulder}
\end{figure}

Our preferred multi-component ice model using  pure and mixed H$_2$O, CO$_2$, and CO ices still cannot fully reproduce the blue CO$_2$ shoulder observed around 4.21--4.23 $\mu$m.  We have investigated a number of possible explanations for this feature, and ruled out that either a higher-temperature (70 K) CO$_2$ component or a high-ratio (1:2) CO$_2$:H$_2$O mixture is responsible.  On the other hand, we can reproduce the blue wing with the addition of a warmed CO$_2$:CO ice component (Figure \ref{fig:CO2_shoulder}).  We find that a 1:1 CO$_2$:CO ice warmed to 70 K and accounting for $\sim$13\% of the total CO$_2$ abundance can very well reproduce the blue CO$_2$ shoulder.  

This blue wing has been previously observed in laboratory spectra of mixed CO$_2$:CO ices heated above $\sim$50 K \citep{vanBroekhuizen2006}.  In our experiments, there is virtually no CO left in the ice at these temperatures, leading us to speculate that the blue-shifted CO$_2$ band is due to the loss of CO from the CO$_2$:CO mixture, leading to a highly porous CO$_2$ structure.  Indeed, absorption bands around these wavelengths have been seen in matrix isolation studies of CO$_2$ clusters \citep{Knoezinger1995}, which could be analogous to very porous CO$_2$ ices with few interactions between neighboring molecules.  If responsible for the blue shoulder of the CO$_2$ feature towards HH 48 NE, this would indicate a component of initially CO-rich CO$_2$ ice that has undergone thermal processing.  

Alternatively, it is also possible that the blue shoulder represents a CO$_2$ component formed from radiation processing of other C- and O-bearing molecules, which is known from lab experiments to produce a blue-shifted band profile \citep[e.g.][]{Ioppolo2009, Jones2014}.  The shoulder could also be a radiative transfer effect.  \citet{Dartois2022} demonstrated the sensitivity of the CO$_2$ band profile to grain and ice compositions and disk viewing geometries for a generic protoplanetary disk model.  At least one scenario ($\sim$73$^\circ$ inclination, ice model 13) can produce a blue absorption excess and red absorption deficit, including a pronounced blue absorption shoulder as seen in HH 48 NE.  Obtaining absorption and scattering coefficients using a more sophisticated treatment of grain agglomeration is required to determine if this is a plausible explanation for the blue CO$_2$ shoulder in HH 48 NE.  

\FloatBarrier
\section{Modeled ice abundances}
\label{sec:app_abund}

Table \ref{tab:abundances} lists the abundances of each ice component for the different models presented in this paper.  M1 is the preferred model (Section \ref{subsec:res_finetune}).  As noted in Section \ref{sec:discussion}, the absolute abundances should be treated with caution due to uncertainties in the model physical parameters like dust settling and grain size distribution.  The abundance ratios are less sensitive to these choices.

We estimate the uncertainties on the M1 model abundance ratios based on two potential sources of uncertainty.  First, we tested the sensitivity to optical depth uncertainties introduced by baseline subtraction (Section \ref{subsec:method_compare}).  This is done by determining the change in input abundance required to produce a given change in observed optical depth.  For the H$_2$O, CO$_2$, and CO bands, the optical depth uncertainties of $\sim$15, 5, and 10\% translate to abundance uncertainties of $\sim$33, 11, and 17\%, respectively.  Second, we compared the abundance ratios inferred from the two most plausible ice mixing scenarios, S3 and M1.  For H$_2$O and CO$_2$ there is percent-level variation between these scenarios, while for CO there is a $\sim$19\% difference.  Taken together, we estimate the uncertainties on the CO$_2$/H$_2$O and CO/H$_2$O abundance ratios from model M1 to be 35\% and 41\%, respectively.  Quantifying any additional error due to varying responses of the H$_2$O, CO$_2$, and CO ice bands to the adopted model physical properties will require a more sophisticated fitting approach.

\FloatBarrier

\begin{deluxetable*}{lcc}
	\tabletypesize{\footnotesize}
	\tablecaption{Modeled ice abundances for different mixing scenarios \label{tab:abundances}}
	\tablecolumns{3} 
	\tablewidth{\textwidth} 
 	\tablehead{
        \colhead{Model}       & 
        \colhead{Zone}        &
        \colhead{Ice abundances $\mathrm{n_X/n_H}$} 
        }
\startdata
\multicolumn{3}{c}{\textit{Single-component models}} \\
\hline 
S1. Entirely pure & Zone 1 & H$_2$O (4.0 $\times$10$^{-4}$) \\
& Zone 2 & Zone 1 + CO$_2$ (4.0$\times$10$^{-5}$) \\ 
& Zone 3 & Zone 2 + CO (2.9$\times$10$^{-5}$)\\
\hline 
S2. Entirely mixed & Zone 1 & H$_2$O (1.35$\times$10$^{-4}$) + H$_2$O:CO$_2$ 10:2 (2.4$\times$10$^{-4}$) + H$_2$O:CO 10:2 (1.2$\times$10$^{-4}$) \\
& Zone 2 & Zone 1 \\ 
& Zone 3 & Zone 1 \\
\hline 
S3. Separate polar/apolar & Zone 1 & H$_2$O (3.9$\times$10$^{-4}$) \\
& Zone 2 & Zone 1 + CO$_2$:CO 3.5:1 (4.7$\times$10$^{-5}$) \\ 
& Zone 3 & Zone 2 \\
\hline 
\multicolumn{3}{c}{\textit{Multi-component models}} \\
\hline 
M1. Preferred & Zone 1 & H$_2$O (2.7$\times$10$^{-4}$) + H$_2$O:CO$_2$ 10:1 (7.5$\times$10$^{-5}$) + H$_2$O:CO 10:1 (4.6$\times$10$^{-5}$)\\
(Polar 10:1, apolar 3.5:1) & Zone 2 & Zone 1 + CO$_2$:CO 3.5:1 (3.75$\times$10$^{-5}$) \\ 
& Zone 3 & Zone 2 \\
\hline
M2. Polar 10:2 & Zone 1 & H$_2$O (3.3$\times$10$^{-4}$) + H$_2$O:CO$_2$ 10:2 (4.3$\times$10$^{-5}$) + H$_2$O:CO 10:2 (2.3$\times$10$^{-5}$)\\
& Zone 2 & Zone 1 + CO$_2$:CO 3.5:1 (3.75$\times$10$^{-5}$) \\ 
& Zone 3 & Zone 2 \\
\hline
M3. Polar 10:3 & Zone 1 & H$_2$O (3.5$\times$10$^{-4}$) + H$_2$O:CO$_2$ 10:3 (3.8$\times$10$^{-5}$) + H$_2$O:CO 10:3 (1.5$\times$10$^{-5}$)\\
& Zone 2 & Zone 1 + CO$_2$:CO 3.5:1 (3.75$\times$10$^{-5}$) \\ 
& Zone 3 & Zone 2 \\
\hline
M4. Apolar CO$_2$ + CO & Zone 1 & H$_2$O (2.9$\times$10$^{-4}$) + H$_2$O:CO$_2$ 10:1 (7.5$\times$10$^{-5}$) + H$_2$O:CO 10:1 (4.6$\times$10$^{-5}$)\\
& Zone 2 & Zone 1 + CO$_2$ (3.4$\times$10$^{-5}$) \\ 
& Zone 3 & Zone 2 + CO (1.9$\times$10$^{-5}$) \\
\hline 
\multicolumn{3}{c}{\textit{Variations on Preferred model M1}} \\
\hline 
V1. Polar CH$_3$OH & Zone 1 & H$_2$O (4.1$\times$10$^{-4}$) \\
& Zone 2 & Zone 1 + CO$_2$:CO 3.5:1 (3.75$\times$10$^{-5}$) + CO$_2$:CH$_3$OH 1:1 (2.1$\times$10$^{-5}$) \\ 
& Zone 3 & Zone 2 + CO:CH$_3$OH 1:1 (4.3$\times$10$^{-5}$) \\
\hline 
V2. CO$_2$ shoulder & Zone 1 & H$_2$O (2.7$\times$10$^{-4}$) + H$_2$O:CO$_2$ 10:1 (7.5$\times$10$^{-5}$) + H$_2$O:CO 10:1 (4.6$\times$10$^{-5}$)\\
 & Zone 2 & Zone 1 + CO$_2$:CO 3.5:1 (3.6$\times$10$^{-5}$) + 70 K CO$_2$:CO 1:1 (5.3$\times$10$^{-6}$) \\ 
& Zone 3 & Zone 2 \\
\hline 
V3. No entrapment & Zone 1 & H$_2$O (4.1$\times$10$^{-4}$) \\
& Zone 2 & Zone 1 + CO$_2$ (1.7$\times$10$^{-5}$) \\ 
& Zone 3 & Zone 2 + H$_2$O:CO 10:1 (1.3$\times$10$^{-4}$) + CO$_2$:CO 3.5:1 (3.8$\times$10$^{-4}$)\\
\hline 
\enddata
\tablenotetext{}{Colons (e.g.~H$_2$O:CO) represent interacting mixtures while pluses (e.g.~H$_2$O + CO) represent non-interacting layers.}

\end{deluxetable*}

\end{document}